\documentclass[%
 reprint,
amsmath,amssymb,
aps,
prl,
twocolumn,
superscriptaddress,
showkeys
]{revtex4-1}

\usepackage[pdftex]{graphicx}
\usepackage{subfigure}
\usepackage{epstopdf}
\usepackage{graphicx}
\usepackage{dcolumn}
\usepackage{bm}
\usepackage[colorlinks,
linkcolor=blue, citecolor=blue, anchorcolor=blue,
urlcolor=blue,
]{hyperref}
\usepackage{braket}
\usepackage{color}
\usepackage{tikz}
\usepackage{textcomp}

\newcommand{\Yb}{$^{171}\textrm{Yb}^+~$}

\newcommand{\Ca}{$^{40}\textrm{Ca}^+~$}

\newcommand{\avg}[1]{\ensuremath{\left\langle#1\right\rangle}}

\usepackage[normalem]{ulem}
\definecolor{jaren}{rgb}{0.8,0.0,0.8}
\newcommand\jarensout{\bgroup\markoverwith{\textcolor{jaren}{\rule[0.5ex]{2pt}{1.0pt}}}\ULon}

\begin{document}

\title{Quantum computation and simulation with vibrational modes of trapped ions}

\author{Wentao Chen}
\email{chen-wt17@mails.tsinghua.edu.cn}
\affiliation{Institute for Interdisciplinary Information Sciences, Tsinghua University, Beijing 100084, China}
\author{Jaren Gan}
\email{jarengan@quantumlah.org}
\affiliation{Centre for Quantum Technologies, National University of Singapore, 3 Science Drive 2, 117543 Singapore, Singapore}
\author{Jing-Ning Zhang}
\email{zhangjn@baqis.ac.cn}
\affiliation{Beijing Academy of Quantum Information Sciences, Beijing 100193, China}
\author{Dzmitry Matuskevich}
\email{phymd@nus.edu.sg}
\affiliation{Centre for Quantum Technologies, National University of Singapore, 3 Science Drive 2, 117543 Singapore, Singapore}
\affiliation{Department of Physics, National University of Singapore, 2 Science Drive 3, 117551 Singapore, Singapore}
\author{Kihwan Kim}
\email{kimkihwan@mail.tsinghua.edu.cn}
\affiliation{Institute for Interdisciplinary Information Sciences, Tsinghua University, Beijing 100084, China}

\date{\today}
\begin{abstract}

Vibrational degrees of freedom in trapped-ion systems have recently been gaining attention as a quantum resource, beyond the role as a mediator for entangling quantum operations on internal degrees of freedom, because of the large available Hilbert space. The vibrational modes can be represented as quantum harmonic oscillators and thus offer a Hilbert space with infinite dimension. Here we review recent theoretical and experimental progress in the coherent manipulation of the vibrational modes, including bosonic encoding schemes in quantum information, reliable and efficient measurement techniques, and quantum operations that allow various quantum simulations and quantum computation algorithms. We describe experiments using the vibrational modes, including the preparation of non-classical states, molecular vibronic sampling, and applications in quantum thermodynamics. We finally discuss the potential prospects and challenges of trapped-ion vibrational-mode quantum information processing.
\end{abstract}



\maketitle


\section{I. Introduction}
The trapped-ion system is a leading platform for the construction of universal quantum computers, offering long coherence time, high-fidelity quantum operations, and reliable state preparation and measurement \cite{leibfried2003quantum,Haffner08Quantum,Wineland08Entangled,Ladd10Quantum,Monroe2013Scaling}. 
Recently, the coherence time of a single ion-qubit has been enhanced from 600 s \cite{wang2017single} to 5400 s \cite{wang2021single}, a universal set of high-fidelity gates reached to 99.999 \% and 99.9 \% for single qubit and two-qubit gates, respectively \cite{Harty2014High,ballance2016high,gaebler2016high}, the number of entangled qubits also has been increased to 24 qubits \cite{Monz2014Qubit,pogorelov2021compact}, and the number of qubits in quantum simulation has been reported to up to 53 qubits \cite{friis2018observation,zhang2017observation}.

While previous research focused mostly on internal state-based quantum information processing, recent experimental advances in the precise control of the motional modes of trapped ions has led to motion-based quantum computation and quantum simulation as well, exploiting an infinite-dimensional Hilbert space and thus potentially yielding greater computational power. Specifically, Fock states of the motional modes with $\sim 10-100$ excitations or more~\cite{McCormick2019Quantum,alonso2016generation}, also called phonons, have been prepared, manipulated, and measured with high fidelity in ion-trap experiments. 
With 3 motional modes and up to 10 phonons each, the dimension of available motional Hilbert space of a single trapped ion is on the order of $10^3$, which is roughly equivalent to that of ten qubits.
It is thus natural to anticipate that one would require a smaller number of ions for the motion-based trapped-ion quantum computer to outperform classical computers in some specialized tasks, compared to its internal state-based counterpart \cite{Paternostro2005Vibrational,Ortiz2016Continuous}.

The idea that the states of the harmonic oscillators can be used for quantum simulation and quantum computations was first proposed in optics~\cite{Braunstein2005Quantum}. The mode of the electromagnetic field can be described as a harmonic oscillator, and it is possible to implement several forms of interaction between the modes with linear optics elements, such as beam-splitters. It was recently suggested~\cite{HKLau2017quantum} that these systems were able to carry out computations for various problems, such as linear component analysis and solving linear equations, and even demonstrating quantum supremacy using the boson sampling approach. However, the technical challenges faced in optical systems, such as photon losses, imperfect photon detectors, non-deterministic single photon generation, and weak nonlinear interaction between light modes makes it difficult to scale up photonic systems \cite{Zhong201812-Photon,Wang2019Boson}. Despite of the difficulty, a Gaussian boson sampling, which might be less problematic for photon loss, with up to 50 indistinguishable single-mode squeezed states into a 100-mode interferometer has been reported \cite{zhong2020quantum}.    

In contrast, trapped ions systems are well developed and widely adopted for the preparation and characterization of non-classical states of harmonic oscillators~\cite{leibfried2003quantum,An2015Experimental,Um16Phonon,Lv2016Reconstruction,Zhang2018Experimental,shen2018quantum}. 
Moreover, it is possible to introduce strong nonlinear interactions, essential for universal quantum computation and quantum simulation~\cite{Lloyd1999Quantum,Braunstein2005Quantum,Zhang2006Continuous,menicucci2006universal}, among vibrational modes of trapped ions through the laser-induced internal state-motion coupling \cite{leibfried2003quantum,Haljan2005Spin-Dependent} as well as via anharmonicity of the Coulomb interaction between trapped ions ~\cite{Ding2014Microwave,Ding2017Quantum,Maslennikov2019quantum,Ding2017Cross}. In conclusion, the above mentioned features have made motional states of trapped ions a highly attractive platform for the construction of a bosonic quantum computer and quantum simulations.

In this article, we review some recent progress in quantum computation and quantum simulation with vibrational modes of trapped ions. In Sec. II, we provide a general framework for the motion-based trapped-ion quantum computer. In Sec III, we introduce the trapped-ion system and the initialization and detection techniques for the motional states of trapped ions. Then we elaborate on Gaussian and non-Gaussian operations on the motional states in Sec. IV and Sec. V, respectively. In Sec. VI, we describe some applications of motion-based trapped-ion quantum computation and quantum simulation. Finally, we conclude the article with some discussion and outlook.

\section{II. Universal quantum computation and simulation with phonons}
Quantum computation and quantum simulation with phonons can be regarded as a physical realization of continuous variable (CV) quantum information processing, which has been developed in various physical platforms such as optical systems \cite{vanLoock2000Multipartite,Knill2001scheme,Kok2007Linear} or microwave modes of electromagnetic radiation and spin ensembles \cite{Hammerer2010Quantum}. Recently, there have been theoretical developments \cite{Weedbrook2012Gaussian} for quantum computation with continuous variables that includes a proposal for universal computing independent of encoding \cite{Lau2016Universal}.   

Similar to the qubit-based circuit model, CV quantum information processing also consists of three essential steps including initialization (or state preparation), unitary manipulation (or operation) and detection (or measurement) of the final quantum states \cite{Braunstein2005Quantum}. It has been clearly shown that in the CV model, it is necessary to include any form of nonlinearity to perform universal quantum computation or non-trivial quantum simulation, which cannot be simply computed by classical computers~\cite{Lloyd1999Quantum}. The nonlinearity can arise from the choice of initial states, number resolving measurements, or non-Gaussian operations. 

Nonlinear states refer to Fock states, or certain superpositions thereof, that cannot be generated by Gaussian operations. Gaussian operations, such as displacement and squeezing operations, are unable to provide the necessary nonlinearity. Single phonon detection and number resolving detection can also introduce the necessary nonlinearity in the protocol. Additionally, non-Gaussian operations can be achieved either by coupling motional modes to the internal spin state of the ion, or by implementing nonlinear interactions among motional modes. We will discuss these requirements in details in the following chapters.  


\section{III. Encoding, initialization and detection of phonon number states}
\subsection{A trapped ion system}

\begin{figure}[tb]
\centering
\includegraphics[width=\columnwidth]{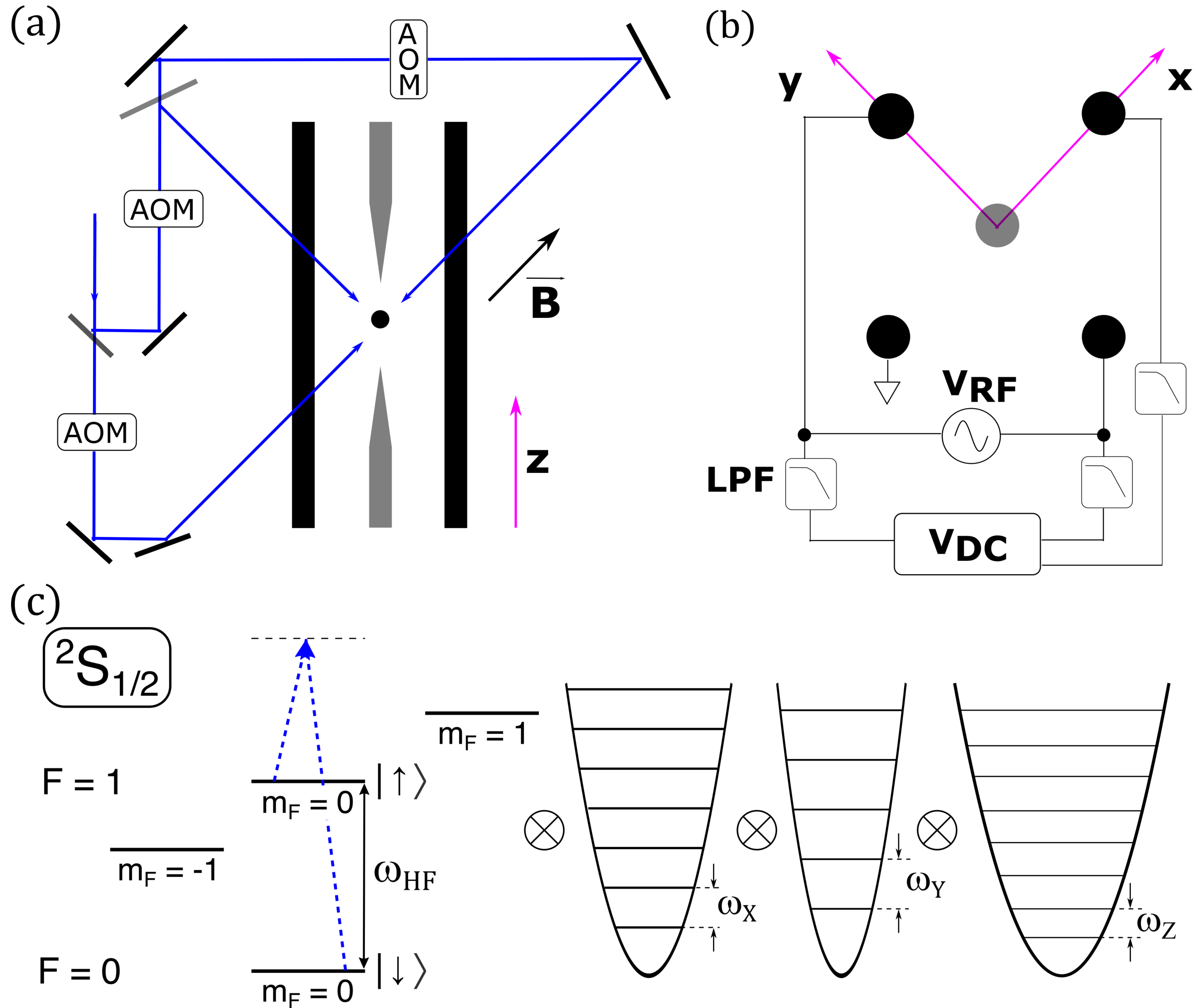}
\caption{\label{fig:IonSetup}
Raman scheme and level diagram of $^{171}\mathrm{Yb}^+$. 
(a) Raman transitions are carried out via laser beams (blue arrows) directed to the ion (filled circle) after passing through acousto-optical modulators (AOMs), which are responsible for controlling the frequency difference between pairs of beams. The magnetic field is directed along a pair of counter-propagating beams, and various combination of polarization for each Raman beam are allowed in this configuration. 
(b) In addition to an rf source driving two diagonally opposite rods, dc voltages are also applied to three of the linear Paul rf trap rods. This allows further control of the radial mode frequencies, $\omega_{\rm X}$ and $\omega_{\rm Y}$. LPF: low-pass filter.
(c) Raman transitions (blue dashed arrows) are carried out between two levels in the $^{2}\mathrm{S}_{1/2}$ manifold of $\mathrm{^{171}Yb^+}$ with a frequency splitting of $\omega_{\rm HF}$, denoted as $\ket{\downarrow} \equiv \ket{\mathrm{F=0,m_F=0}}$ and $\ket{\uparrow} \equiv \ket{\mathrm{F=1,m_F=0}}$. The motional modes X, Y, and Z with corresponding mode frequencies $\omega_{\rm X}$, $\omega_{\rm Y}$, and $\omega_{\rm Z}$ can also be involved in the Raman process by controlling the frequency difference between two raman beams. In this manner, single and two mode quantum operations can be implemented. Adapted from Ref. \cite{Ding2017Quantum}.
}
\end{figure}

In a trapped ion system, a qubit can be realized by using two stable energy levels separated by an optical transition (quadrupole or octupole transition), a hyperfine transition, or a Zeeman transition in an atomic ion. In this review, the atomic ion species $\mathrm{^{171}Yb^+}$ is used as a concrete example, but all the discussion here can be directly applied to any other ion-qubits. The two levels in the hyperfine structure of $^{2}\mathrm{S}_{1/2}$ manifold are usually used to realize a qubit, which are denoted as $\ket{\downarrow} \equiv \ket{\mathrm{F=0,m_F=0}}$ and $\ket{\uparrow} \equiv \ket{\mathrm{F=1,m_F=0}}$ with a frequency difference of $\omega_{\rm HF}=(2\pi)12.6428$ GHz. Typically, the ion is trapped by a radio-frequency Paul trap, which provides a harmonic trapping potential with a trap frequency $\omega_{m} ( = \{\omega_{\rm{X}},\omega_{\rm{Y}},\omega_{\rm{Z}}\})$ in the MHz range. 
The state of the system is represented in Fock state basis as $\Ket{\sigma,n_{m}}$, where $\sigma$ is the state of the qubit and $n_{m}$ is the phonon number in the vibrational mode. Two laser beams from a pico-second pulsed laser detuned from the $^{2}\mathrm{S}_{1/2}\rightarrow ^2\mathrm{P}_{1/2}$ transition are used to generate a stimulated Raman process to drive the carrier and motional sideband transitions of the ion \cite{hayes2010entanglement} when the beatnote frequency between Raman laser beams is resonant with the qubit and sideband frequencies, respectively. These transitions are described by the time evolution of the following interacting Hamiltonians $H_{{\rm c}}$ and $H_{m}$, respectively \cite{leibfried2003quantum}: 
\begin{eqnarray}
H_{{\rm c}}&=&\frac{\Omega_{{\rm c}}}{2}\left(\sigma_{+}e^{i\varphi_{\mathrm{c}}}+\sigma_{-}e^{-i\varphi_{\mathrm{c}}}\right), \label{eq:Car}\\ H_{{\rm b(r)}, m}&=&\frac{i\eta_{m}\Omega_{m}}{2}\left(\sigma_{\pm}a_{m}^{\dagger}-\sigma_{\mp}a_{m}\right),\label{eq:RsbBsb}
\end{eqnarray}
where $\Omega_{{\rm c}}$ and $\eta_{m}\Omega_{m}$ are the Rabi frequencies of carrier and blue (red) sideband transitions, $\sigma_{+}=\Ket{\uparrow}\Bra{\downarrow}$ and $\sigma_{-}=\Ket{\downarrow}\Bra{\uparrow}$ are the spin ladder operators, and $a_{m}^{\dagger}$ ($a_{m}$) is the creation (annihilation) operator of the motional mode $m$. By tuning the beatnote frequencies to address the 2nd order sidebands, a two-phonon creation (annihilation) operation with internal state exchange can be realized,
\begin{eqnarray}
H_{{\rm b2(r2)},m}&=&\frac{i\eta^2_{m}\Omega_{m}}{2}\left(\sigma_{\pm}a_{m}^{\dagger 2}-\sigma_{\mp}a^
{2}_{m}\right).\label{eq:RSBBsb2}
\end{eqnarray}

\subsection{Ground-state cooling} 

Generally, manipulation of quantum states of motion begins with preparing all the relevant modes in the ground state.
Initialization to the ground state is usually carried out with a series of laser cooling techniques. 
These cooling methods provide a practical way of directly reducing entropy of the trapped ion system, by transferring energy from the motion to the scattered photons. 
Laser cooling was first demonstrated using a velocity-dependent radiative force \cite{wineland1978radiation,neuhauser1978optical}, called Doppler cooling. The final temperature that can be achieved with Doppler cooling is limited by the natural linewidth of the transition involved. 
Typically, after Doppler cooling, the mean phonon number reaches roughly a few tens in typical trap frequencies of a few MHz. 
Sisyphus coolings \cite{dalibard1989laser,ejtemaee20173d} and EIT coolings \cite{morigi2000ground,roos2000experimental,lin2013sympathetic,lechner2016electromagnetically,jordan2019near} can further lower the temperature, going below the Doppler limit. 
For a single ion, mean phonon numbers between 1 and 0.1 have been reported after Sisyphus cooling and EIT cooling, respectively. 
The final step to achieving ground state is then realized by  resolved-sideband cooling~\cite{monroe1995resolved, roos1999quantum}. 

\subsection{Encoding scheme of phonons and initialization}
There are various initial phonon states depending on the types of quantum simulation and quantum computation. For boson sampling and Gaussian boson-sampling experiments, the initial state can be a Fock state, a coherent state, a squeezed vacuum state, displaced squeezed state, or any combination of them. For the machine learning algorithm, a classical vector ${\alpha_n}$ of $n$ complex numbers is mapped to the quantum state $\sum_n \alpha_n |n\rangle$, where $|n\rangle$ denotes the Fock state. For quantum computation with bosonic modes, a Schr\"{o}dinger Cat state \cite{monroe1996schrodinger}, a binomial state \cite{michael2016new}, and  Gottesman, Kitaev and Preskill (GKP) states~\cite{gottesman2001encoding} have been used for logical qubits.  

Most of these initial states can be implemented by applying either a coherent laser beam, or Raman laser beams on the ions, which creates a Jaynes-Cummings type of coupling (Eq.~\ref{eq:RsbBsb}) that implements displacement and squeezing operations. The details of these operations are elaborated in the next section. 
An excited Fock state ($n \neq 0$) and an arbitrary superposition of Fock state with phonons can be efficiently generated by the scheme demonstrated in the Be$^+$ ion system~\cite{Ben2003Experimental}. 
With the assistance of the ion's internal state, an arbitrary quantum superposition of Fock states can thus be created using interactions of the Jaynes-Cummings type. 
By applying the displacement operation on the vacuum state, a coherent state can be obtained~\cite{Meekhof1996Generation}. 
The Schr\"{o}dinger Cat state can be generated by an internal state dependant displacement operation \cite{monroe1996schrodinger}, typically known as the spin-dependant force \cite{Haljan2005Spin-Dependent,lee2005phase}. 
Additionally, the squeezing operation generates squeezed vacuum-state and displaced-squeezed state if it is applied to a vacuum, or coherent state \cite{Meekhof1996Generation,leibfried2003quantum,lo2015spin}, respectively. 
However, we note that these initial states can also be generated by specially designed bath-engineering~\cite{kienzler2015quantum}.   

\subsection{Detection of phonon states}
In an optical system, photon detectors or photon counters are used to measure photon states. Based on the measured photon results, various tomography of photon states including Winger function, $Q$-function, $P$-function, and Fock-state tomography are performed. However, for phonons in a trapped ion system, such phonon detectors or counters do not exist. 
The phonon number states, and hence its population distribution, are detected through their interaction with the internal states of the ions.
Then, similar to the optical system, the measured phonon-number states can be used to reconstruct the phonon density matrix, Wigner-functions, and $Q$-functions. The detection methods of phonon numbers can be categorized as follows:
  
- {\it Phonon state detection by interference:} The traditional phonon state detection is performed by observing the phonon-number dependent Rabi oscillations, typically by driving the blue-sideband transition. More precisely, the probability of occupying a phonon number state $\ket{n}$,  $P_{n}$, can be determined from a Fourier transform of the time evolution of $P_{\uparrow} (t)$, the probability of detecting the ion in the $\ket{\uparrow}$ state~\cite{Meekhof1996Generation}:
\begin{eqnarray}
P_{\uparrow} \left(t\right) =\frac{1}{2} \sum_{n} P_{n} \left[1- e^{-\gamma_{n} t}A\cos\left(2 \Omega_{n,n+1} t\right)\right].
\label{eq:bsbfit}
\end{eqnarray}
Here $P_{\uparrow}(t)$ is experimentally measured with state dependent fluorescence detection~\cite{Olmschenk07Manipulation,Zhang12State}. The decay rate $\gamma_{n}$ factors in fluctuations of laser intensity, and as well heating of the motional mode. The dependence on $n$ can be empirically determined~\cite{Meekhof1996Generation}. The contrast $A$ is required to take into account the imperfections of state preparation and detection. 
In the Lamb-Dicke regime, $\eta_m n^2 \ll 1$ \cite{leibfried2003quantum}, the approximation of phonon-number dependent Rabi frequencies $\Omega_{n,n+1} \approx \sqrt{n+1} \eta_m\Omega_m$ can be made. 
We note that as this method is not a projection measurement of the phonon state, it thus has limited applicability for the correlation measurement of more than a single mode. 

- {\it Phonon number resolving detection by non-linearity:} Projective measurements of the motional state of ions can be realized through an effective cross-Kerr coupling to a second motional mode. 
Direct single-shot measurements of phonon number states and as well parity of the harmonic oscillator (defined as the difference between the even and odd phonon number state occupations) can be realized by utilizing the nonlinear interaction among motional modes.
This detection scheme involving parity measurements has been used to reconstruct the Wigner function of a motional state~\cite{Ding2017Quantum}. The ability to measure parity of the oscillator is discussed in more details in section IV and Ref. \cite{Ding2017Cross}.

\begin{figure}[tb]
\centering
\includegraphics[width=\columnwidth]{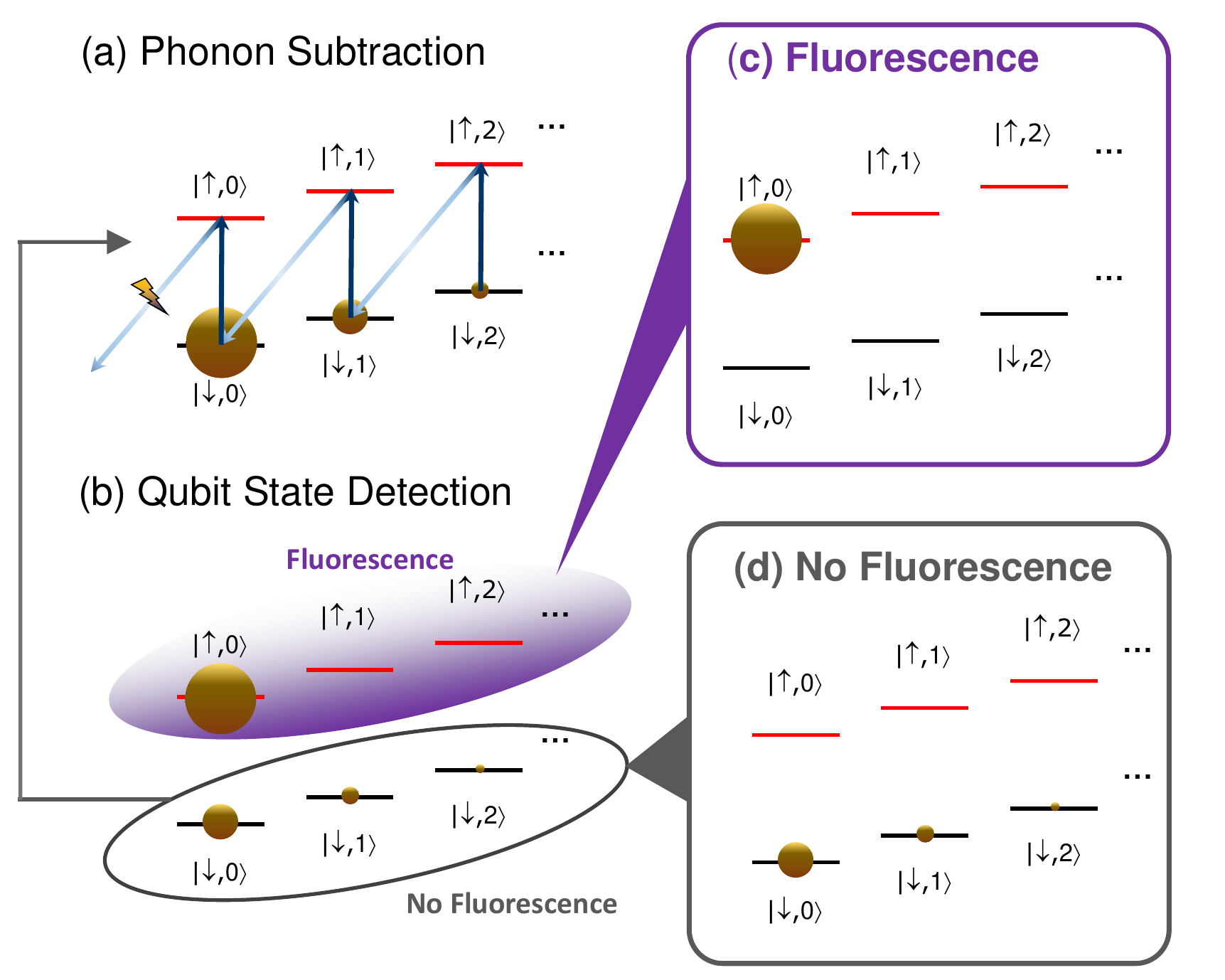}
\caption{\label{fig:ProjectMeas}
(a) The phonon subtraction operation that changes phonon state from $\ket{n}$ to $\ket{n-1}$ is composed of resonant $\pi$-pulses on the carrier transition (black arrow) followed by an adiabatic drive on the blue-sideband (grey arrow). 
The subtraction $\ket{n}\rightarrow\ket{n-1}$ is performed for every Fock state apart from $\ket{n=0}$, which is driven to the state $\ket{\uparrow, n=0}$. (b) On application of the detection laser (c), the fluorescence observed is attributed to the population in the state $\ket{\downarrow, n=0}$ prior to phonon subtraction. (d) No fluorescence is detected from the other Fock states, which have had one quanta removed after the operation, and are shelved to the $\ket{\downarrow}$ state. The initial state of $\ket{\downarrow, n}$ thus arrives at $\ket{\uparrow, n=0}$ after ($n+1$) repetitions of (a)(b) procedures, and is detected via fluorescence. Adapted from Ref.~\cite{An2015Experimental}}
\end{figure}

{\it - Phonon number resolving detection by phonon subtraction operations:} 
By implementing a repeated sequence of phonon subtraction and qubit state detection, the phonon number state can be determined (and projected) by counting the number of repetitions before fluorescence is first observed. Upon detection of fluorescence in the $(n+1)$-th iteration, the phonon number state will have been the Fock state $\ket{n}$.
The phonon subtraction $|n\rangle \rightarrow |n-1\rangle$ is performed by the successive application of a carrier $\pi$-pulse driving $\ket{\downarrow,n} \rightarrow \ket{\uparrow,n}$, and a uniform blue-sideband transition that drives $\ket{\uparrow,n} \rightarrow \ket{\downarrow,n-1}$ as shown in Fig. \ref{fig:ProjectMeas}. 
After a single instance of phonon subtraction, only the state $\ket{\downarrow,0}$ is transferred to the bright $\ket{\uparrow,0}$ state that scatters photons at the qubit-detection step. 
This scheme has been extended to map phonon number states to auxiliary internal states \cite{ohira2019phonon}. 
We note that if the sequence is applied only once, the population of the ground-state $\ket{0}$ is measured, which allows reconstruction of the $Q$-function~\cite{Lv2016Reconstruction}. 
The method of projective measurement can further be extended to enable correlation measurements of multiple modes. For the $N$ motional mode detection, the use of $N$ qubits~\cite{An2015Experimental,Um16Phonon,Lv2016Reconstruction} are required. 

\section{IV. Gaussian operations}


\begin{figure*}[htb]
\includegraphics[width=0.9\textwidth]{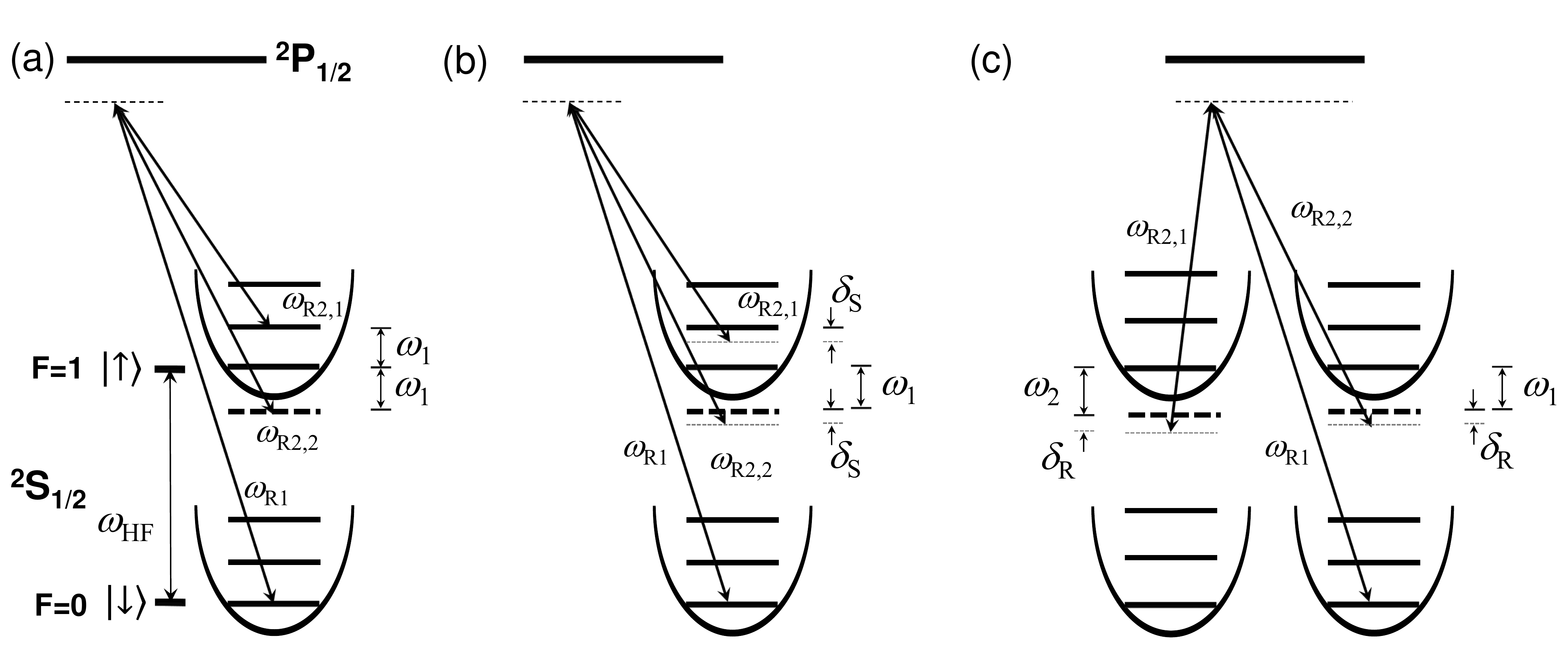}
\caption{\label{FigureRSD} Trapped-ion implementation of the quantum optical operations. The Hilbert space is composed of two phonon-modes, 1 and 2, and the internal electronic state, $\ket{\uparrow}$ and $\ket{\downarrow}$. The quantum operations are implemented via control of the frequencies $\omega_{\rm R2,1}$, $\omega_{\rm R2,2}$, and phases $\phi_1$, $\phi_2$. (a) Coherent displacement operation $\hat{D}$ and (b) squeezing operation $\hat{S}$ on the mode 1 as an example. (c) Rotation operation $\hat{R}$ between mode 1 and mode 2.
Adapted from Ref. \cite{shen2018quantum}}.
\end{figure*}


A Gaussian operation transforms Gaussian states into Gaussian states. Typical examples of Gaussian operations are displacement and squeezing operations. There exist many schemes to realize these Gaussian operations for vibrational modes of trapped ions \cite{leibfried2003quantum,monroe1996schrodinger,Meekhof1996Generation,leibfried1996experimental}. For example, a displacement operation can be implemented by applying an electric field for a certain duration in a desired direction. A squeezing operation can be realized by modulating the trapping potential at twice the trap frequency, also known as a parametric drive. These Gaussian operations can also be realized by applying two Raman laser beams with suitable polarization and frequency difference between them. The implementation of displacement or squeezing operations can be state-independent if the frequency differences are $\omega_m$ or $2\omega_m$, similar to the electric field displacement or parametric drive~\cite{leibfried2003quantum,ge2019trapped}. However, these state-independent operations are challenging to incorporate together with blue- or red-sideband transitions, when well-defined phase relations between them are required. Therefore, the state-dependant operations are more appropriate for the general manipulation of phonons for quantum computation and quantum simulations. 

The state-dependant Gaussian operations require one frequency in a laser denoted Raman 1, $\omega_{\rm R1}$, and two frequencies and phases in the other laser denoted Raman 2, $\omega_{\rm R2,1}$ and $\omega_{\rm R2,2}$, where it is assumed that Raman 1 and Raman 2 are propagating towards the ion from two different directions (Fig.~\ref{fig:IonSetup}a). 
The Hamiltonian of an ion in a harmonic oscillator with the Raman laser beams can be expressed with the following equation,
\begin{align}
H&=\dfrac{\hbar \omega_{\rm HF}}{2} \sigma_{\rm z} + \hbar \omega_{m} (a_{m}^\dagger a_{m}+\dfrac{1}{2}) \nonumber\\
&+  \sum_{j=1,2}\dfrac{\hbar \Omega}{2}(\sigma_+ + \sigma_-)(e^{i (\omega_{L,j} t+\phi_{j})} + e^{-i (\omega_{L,j} t+\phi_{j})}),
\label{LightmatterHam}
\end{align}
where $\sigma_{\rm z}$ is the Pauli matrix, $\Omega$ is the effective Rabi frequency, $\omega_{L,j}=\omega_{R1}-\omega_{R2,j}$ are the effective laser frequencies, and $\phi_{j}$ are the phases.

The interaction Hamiltonian with respect to $H_0=\dfrac{\hbar \omega_{\rm HF}}{2} \sigma_{\rm z} + \hbar \omega_{m} (a_{m}^\dagger a_{m}+\dfrac{1}{2})$ after applying the rotating wave and Lamb-Dicke approximations can be written as
\begin{align}
H_{I} &= \sum_{j=1,2} \dfrac{\hbar \Omega}{2}\sigma_+  \{1+ i\eta_{m}(a_{m} e^{-i\omega_{m}t}+a_{m}^\dag e^{i\omega_{m}t}) \nonumber\\&- \eta_{m}^2(a_{m} e^{-i\omega_{m}t}+a_{m}^\dag e^{i\omega_{m}t})^2\} e^{-i \delta_{j} t} e^{i\phi_{j}} +{\rm h.c.},
\end{align}
where $\delta_{j}=\omega_{L,j}-\omega_{\rm HF}$.

When we consider the resonant terms, we have the effective Hamiltonian of a displacement operation $H_{\rm D} = (1/2) \hbar \eta_{m} \Omega \left( a_{m}^\dag e^{i\phi_{\rm B}} + a_{m} e^{-i\phi_{\rm B}}\right) \sigma_{\rm x} $ by setting $\delta_1=\omega_{m}$, $\delta_2=-\omega_{m}$, and $\phi_1=\phi_2=\pi/2$, as shown in Fig.~\ref{FigureRSD}a. The corresponding time evolution of the displacement operation $\hat{D}$ of a single mode can be realized as
\begin{align}
\hat{D}(\alpha_{m}) =\exp\left\{ -i \alpha_{m} ( a_{m}^\dag e^{i\phi_{\rm B}} +a_{m} e^{-i\phi_{\rm B}}) \sigma_{\rm x} \right\},
\end{align}
where $\alpha_{m}= t (1/2) \hbar \eta_{m} \Omega$ and $\phi_{\rm B}=\phi_2-\phi_1=0$. This is also known as the $\sigma_{\rm x}-$dependent displacement operation \cite{Haljan2005Spin-Dependent,lee2005phase}. 
The $\sigma_{\rm z}-$dependent displacement operation can also be realized with additional $\pi/2$ carrier rotation pulses (along $\sigma_{\rm y}$ and $\sigma_{\rm -y}$ axis) before and after the $\sigma_{\rm x}-$dependent displacement operation. 
By preparing the qubit in an eigenstate ($
\sigma_{\rm x}$), the displacement operation can be implemented. 

Similarly, by setting $\delta_1=\omega_{m}-\delta_{\rm S}$, $\delta_2=-\omega_{m}-\delta_{\rm S}$, as shown in Fig. \ref{FigureRSD}b, the squeezing operation $\hat{S}$ of a single mode can be realized as
\begin{align}
\hat{S}(\zeta_{m}) =\exp \left\{-i \zeta_{m} (a_{m}^{\dagger 2} e^{i\phi_{\rm B}} + a_{m}^2 e^{-i\phi_{\rm B}} )\sigma_{\rm z}\right\},
\end{align}
where $\zeta_{m}=t \dfrac{\hbar \eta_{m}^2 \Omega^2}{8}(\dfrac{1}{\delta_1} - \dfrac{2}{\delta_1-\omega_{m}}+\dfrac{1}{\delta_1-2\omega_{m}} )$. Typically $\delta_{\rm S}$ is chosen to be few times that of $\eta_{m} \Omega$. Similarly, the squeezing operation can be implemented by preparing the qubit in an eigenstate ($
\sigma_{\rm z}$).

\subsection{Rotations}

\subsubsection{Rotations of collective modes} 

The rotation of collective modes can be realized by similar Raman processes \cite{shen2018quantum}. By setting $\delta_1=-\omega_1-\delta_{\rm R}$, $\delta_2=-\omega_2-\delta_{\rm R}$ as shown in Fig. \ref{FigureRSD}c, where the magnitude of $\delta_{\rm R}$ is on the order of a few times of $\eta_{1,2} \Omega_{1,2}$, the rotation operation can be written as
\begin{align}
\hat{R}(\theta) = \exp\left\{-i\theta (a_{1}^\dagger a_{2} e^{i\phi_{\rm B}}+a_{1} a_{2}^\dagger e^{-i\phi_{\rm B}} )\sigma_{\rm z}  \right\},\label{eq:BS}
\end{align}
where $\theta = t \eta_1 \eta_2 \Omega_1 \Omega_2 / \Delta_{\rm BS}$. The terms $\eta_i$ and $\Omega_i$ denote the Rabi frequency and Lamb-Dicke parameter of the $i-th$ mode, respectively, and $\dfrac{1}{\Delta_{\rm BS}} = \dfrac{1}{-\delta_1}+ \dfrac{1}{-\delta_1+\omega_{1}-\omega_{2}}+\dfrac{1}{\delta_1-\omega_{1}}+\dfrac{1}{\delta_1+\omega_{2}}$. With the qubit prepared in the ($
\sigma_{\rm z}$) eigenstate, beam-splitting or rotation operations can be implemented. 






\subsubsection{Rotations of local modes}


In a Coulomb crystal made up of ions, each one of them experiences a confining harmonic potential arising from the electrodes, as well as a repulsive potential due to the other ions. The motion of ions at each site can be treated separately from the others, and these are known as local modes~\cite{brown2011coupled,harlander2011trapped}. For a linear ion chain, considering the motions in the transverse direction, the Hamiltonian of such a system can be represented as
\begin{equation}
H_m = \sum_i (\omega_{\rm m}+\nu_i) a_i^{\dagger} a_i + \sum_{i,j} \kappa_{ij} a_i^{\dagger} a_j,     
\end{equation}
where $\omega_m$ denotes the transverse trap frequency, $\nu_i$ is a frequency shift determined by the position of the $i$-th ion, and $\kappa_{ij} = e^2 / (2M \omega_{\rm m} d_{ij}^3)$ describes the hopping rate of phonons \cite{Haze2012Observation,toyoda2015hong} between ion $i$ and $j$ with a separation distance $d_{ij}$. $e$ and $M$ are the charge and mass of a single ion, respectively. To study the interaction between local modes, the time taken for both manipulation and detection operations should be much shorter than the time taken for propagation of mechanical waves between ions. This implies that, when compared to using collective motional modes, ions should be separated by a larger distance, which typically results in a hopping rate of kHz level. 

The hopping of phonons from site to site can be viewed as an energy exchange among local modes, and this effect in turn finds applications in quantum simulation and quantum information processes. It is also possible to generate entangled states between ions in this manner. The first experimental demonstration between local modes was carried out in a linear trap, with two ions held in two separate potential wells~\cite{brown2011coupled,harlander2011trapped}. With precise control of electric potentials, the harmonic frequency of two ions can be made equal, thus eliminating the energy gap between them. The Hamiltonian describing the dipole-dipole interaction between the two ions is
\begin{equation}
\label{eq:dipole-dipole interaction}
\begin{aligned}
U_{\mathrm{dd}} &=-\frac{q_{1} q_{2}}{2 \pi \varepsilon_{0}} \frac{\Delta z_{1} \Delta z_{2}}{r^{3}} \\
&=-\hbar \frac{\Omega_{\mathrm{c}}}{2}\left(a_{1}+a_{1}^{\dagger}\right)\left(a_{2}+a_{2}^{\dagger}\right) \\
& \approx-\hbar \frac{\Omega_{\mathrm{c}}}{2}\left(a_{1} a_{2}^{\dagger}+a_{1}^{\dagger} a_{2}\right),
\end{aligned}
\end{equation}
where $\Omega_{\mathrm{c}}=q_{1} q_{2} / (2 \pi \varepsilon_{0}r^{3} \sqrt{M_{1} M_{2} \omega'_{1} \omega'_{2}})$ denotes the coupling rate, $q_i$ and $M_i$ are the charge and mass of the $i$-th ion, and $r$ is the distance between the two ions. The $a_i (a_i^\dagger)$ terms correspond to the phonon annihilation (creation) operator acting on the $i$-th site, which has an effective local frequency $\omega'_i=\omega_m + \nu_i$. Due to the nature of the interaction (Eq.~\ref{eq:dipole-dipole interaction}), evolution of the phonon dynamics always results in a state that is constrained to a Hilbert space with a constant total phonon number. Using radial modes, similar hopping phenomena has been observed for up to 4 ions in a single trap, where a relatively weak confinement along axial direction was adopted \cite{Haze2012Observation,toyoda2015hong,debnath2018observation,tamura2020quantum,ohira2020confinement}.  

By implementing phonon hopping between the local modes of two ions, the Hong-Ou-Mandel effect for phonons have been observed~\cite{toyoda2015hong}. After cooling to the ground state, two ions prepared in the state $|\downarrow,1\rangle_1 |\downarrow,1\rangle_2$ leads to an almost perfect disappearance of the phonon coincidence, proving the indistinguishability of bosons in the trapped ion system. This method provides a clear way to generate the entangled state $(\ket{2}_1\ket{0}_2 + \ket{0}_1\ket{2}_2 )/ \sqrt{2}$, which was demonstrated in Ref.~\cite{toyoda2015hong} with a fidelity of $0.74\pm0.05$ after correcting for experimental imperfections.

In general, it is challenging to control the operations on local modes in a trapped-ion system. However, similar to a photonic system, a blockade~\cite{debnath2018observation,ohira2020confinement} can be realized by driving a resonant sideband operation that effectively induces a frequency shift to the motional mode. This induced shift in the levels increases the energy cost of phonons occupying the site, and hence decreases the rate of hopping. Phonon blockade can be applied to assist in state preparation of a local mode, or to artificially freeze hopping between specific neighbouring ions in a controlled manner.

\section{V. Non-Gaussian operations} 

To achieve universal quantum computation and simulation, it is necessary to implement non-Gaussian operations. Such operations typically arise from nonlinear interactions in the system, either directly between motional modes, or by coupling motion with the ions' internal spin degree of freedom. With non-Gaussian operations, one would be able to realize arbitrary Hamiltonian, and hence achieve universal quantum computation and simulation. In the following section, we review and discuss various experimental methods and techniques for implementing non-Gaussian operations that are available to the trapped ion system.

\subsection{Nonlinear coupling between qubit and harmonic oscillators}
\subsubsection{Controlled and exponential SWAP operations}

An experimental demonstration of a non-Gaussian operation resulting from nonlinear interactions between spin and motion was realized recently~\cite{Gan2020hybrid}, where a controlled beam splitter (CBS) gate was achieved. Depending on the internal spin state of the ion, the beam splitter operation acting on two motional modes $a_i$ and $a_j$ is implemented according to the Hamiltonian:
\begin{equation}
\label{eq:CBS_H}
H_\textrm{CBS} =  \hbar \xi \ket{e}\bra{e}( a_i^{\dagger} a_j\operatorname{e}^{i\upsilon} + a_i a_j^{\dagger}\operatorname{e}^{-i \upsilon}),
\end{equation}
where the beam splitter operator $a_i^\dagger a_j + h.c.$ is applied with a coupling strength $\xi$ when the internal state is $\ket{e} \equiv \ket{F=1,m_F = 1}$. The phase factor $\upsilon$ is set by system parameters, and can be chosen such that $\upsilon = 0$. This gives rise to the unitary transformation $U(t)=\mathrm{exp}(-\mathrm{i} \xi t \ket{e}\bra{e}( a_i^{\dagger} a_j + a_i a_j^{\dagger}))$.
The CBS gate is realized by applying the transformation for a duration $\tau = \frac{\pi}{2\xi}$, i.e. $U(\tau) = U_\textrm{CBS}$, which transforms Fock states as
\begin{align}
\label{eq:CBS_U}
U_\textrm{CBS}\ket{\downarrow,n,m} = & \ket{\downarrow,n,m}, \nonumber \\
U_\textrm{CBS}\ket{e,n,m} = & (-i)^{n + m}\ket{e,m,n},
\end{align}
where $\ket{n}$ and $\ket{m}$ correspond to Fock states of modes $a_i$ and $a_j$, respectively. The CBS is similar to the well-studied controlled swap gate (CSWAP), which is used in a number of applications including purity measurement, overlap test~\cite{Radim2002overlap}, quantum fingerprinting~\cite{Buhrman2001quantum}, and quantum machine learning~\cite{HKLau2017quantum}. With an additional ancillary mode $c$ prepared in the vacuum state, and judicious choices of the phase $\upsilon$, the CSWAP gate can be obtained from a combination of the conditional beam splitter operation~\cite{GanThesis}
\begin{equation}
\label{eq:CBS_CSWAP}
U_\textrm{CSWAP} = U^{a_i a_j,~\upsilon=\pi/2}_\textrm{CBS}  (U^{a_i c~\upsilon=0}_\textrm{CBS})^2,
\end{equation}
where the superscript denotes the modes coupled by the beam splitter. The term $(U^{a_i c~\upsilon=0}_\textrm{CBS})^2$ that couples to the ancillary mode $c$ applies a parity operation, which is used to cancel out the state-dependent phase factor resulting from the coupling between the modes of interest.

The Hamiltonian (\ref{eq:CBS_H}) is implemented by means of a modulating state-dependent optical dipole force, where the modulation frequency $\omega_L$ exactly matches the frequency difference between the modes $a_i$ and $a_j$. Interfering a pair of laser beams that have mutually orthogonal linear polarization and a frequency difference $\omega_L$ at the position of the ion generates an optical lattice with a resultant polarization that oscillates from left to right-handed circular with a frequency $\omega_L$. The state $\ket{e}$ experiences a modulating ac Stark shift, which results in an optical dipole force modulated with frequency $\omega_L$, while the state $\ket{\downarrow}$ does not~\cite{Wineland1998experimental}. The choice $\omega_L = \vert \omega_{a_i} - \omega_{a_j} \vert$ realizes Hamiltonian (\ref{eq:CBS_H}).

\begin{figure}
\centering
\includegraphics[width=\columnwidth]{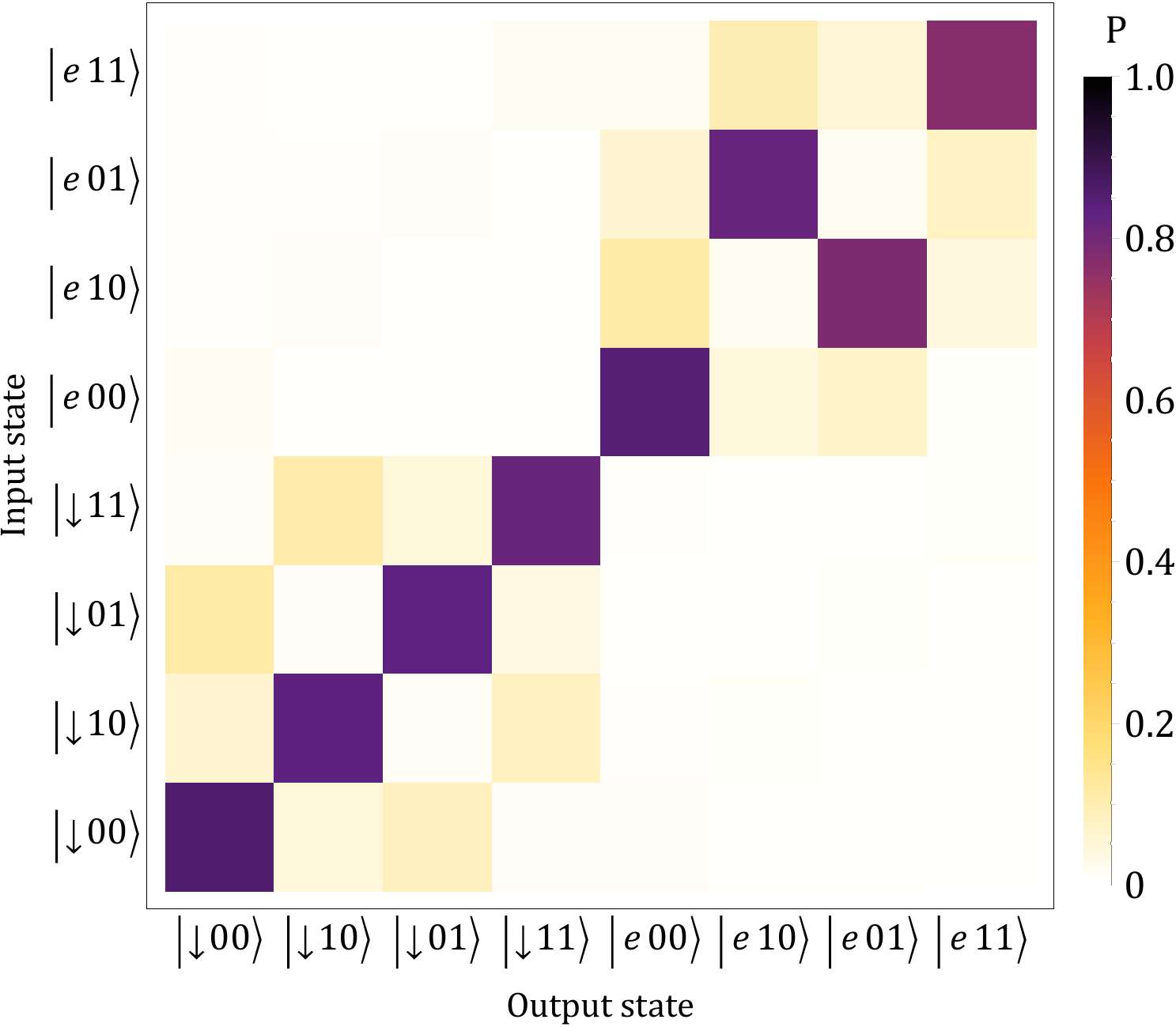}
\caption{
\label{fig:fredkin_table}
Truth table of Fredkin gate. By limiting two modes of the ion to a single phonon each, the controlled beam splitter (CBS) is able to simulate a quantum Fredkin gate. For each of the eight basis states prepared as input, the probabilities of all outcomes after applying the CBS gate are measured via a series of projective measurements. Each data square represents an average of 10000 experiments. Without correcting for state preparation and measurement errors, an average gate success probability of $0.82\pm\,0.01$ is obtained. Adapted from Ref.~\cite{Gan2020hybrid}
}
\end{figure}

The Fredkin gate~\cite{Fredkin1982Conservative,Patel2016quantum, Ono2017Implementation,Linke2018Measuring,Gao2019Entanglement,Zhang2019Modular} is a universal 3-bit gate which swaps the values of the last two bits depending on the first. It can be implemented with a single application of the CBS gate by constraining the accessible Fock states of both motional modes to either $\ket{0}$ or $\ket{1}$. Experimentally, the results are shown in Fig.~\ref{fig:fredkin_table}. To determine the output of the gate after applying it to an input state, projective measurements are used to obtain the probability of each basis state. These measurements are repeated for each basis state as an input. An average of 10000 projective measurements are carried out for each data square, and without correcting for state preparation and measurement errors, an average gate success probability of $0.82\pm0.01$ is obtained.

In addition to the Fredkin gate, the CBS gate also sees applications in other scenarios~\cite{Gan2020hybrid}. Single-shot parity measurements~\cite{Ding2017Quantum}, and hence measurements of Wigner functions, can be carried out by applying the parity operation in Eq.~(\ref{eq:CBS_CSWAP}). When coupled with qubit rotation gates, the CBS gate enables SWAP test measurements which efficiently measures the overlap between two quantum state. Last but not least, a combination of qubit rotation and CSWAP gates allows one to implement the exponential SWAP gate to generate arbitrary entanglement between two modes.

\subsubsection{Phonon arithmetic}
Another type of nonlinear operation can be realized by a simple conventional arithmetic operation, which can be used for phonon number resolving detection. The conventional addition and subtraction of a particle can be written as
\begin{eqnarray}
\hat{S}^{+}=\sum_{n=0} \ket{n+1}\bra{n}, \hat{S}^{-}=\sum_{n=1} \ket{n-1}\bra{n}.
\label{eq:addsubtract}
\end{eqnarray}
where $\ket{n}$ stands for a Fock state of $n$ bosons. $\hat{S}^+$ takes the $n$-particle state to the $(n+1)$-state representing an addition, while for the subtraction operation, $\hat{S}^-$ brings $n$ to the $(n-1)$-state. These operations correspond to conventional arithmetic which is commonly used in everyday life, but they do not emerge naturally in quantum mechanics. Instead, in quantum-mechanics, creation $\hat{a}^\dag$ and annihilation $\hat{a}$ operators are introduced, and they bear the expressions $
\hat{a}^\dag=\sum_{n=0} \sqrt{n+1}\ket{n+1}\bra{n}$ and $\hat{a}=\sum_{n=0}\sqrt{n} \ket{n-1}\bra{n}$, where the proportionality factors $\sqrt{n+1}$ and $\sqrt{n}$ appear due to the symmetric indistinguishable nature of bosons \cite{Dirac1957Principles}. The arithmetic operations, addition and subtraction, in quantum domain thus involves modification of the probability amplitude due to the $n$-dependent factor. 

\begin{figure}
\centering
\includegraphics[width=\columnwidth]{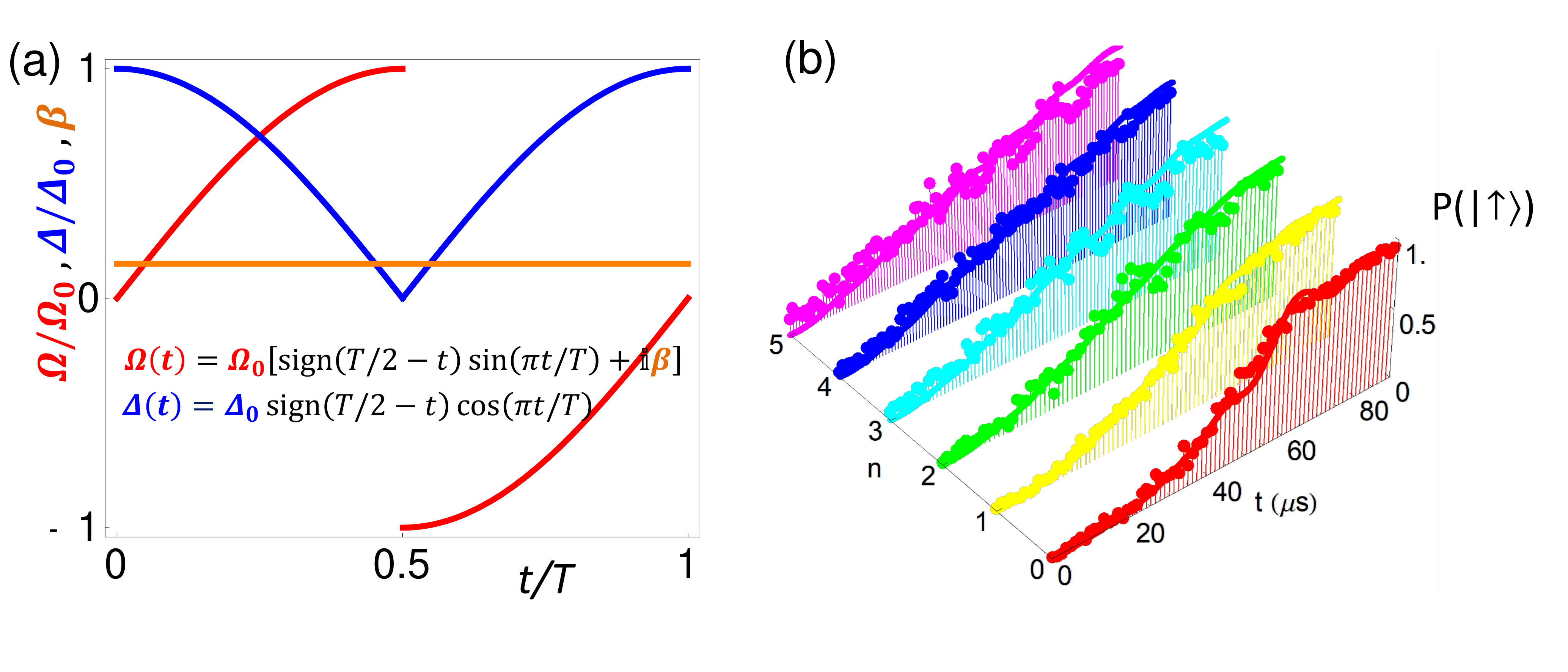}
\caption{
\label{fig:UniformBSB}
(a) For the uniform blue-sideband transition whose frequency is independent of motional quantum number $n$, $\Omega(t)$ (Rabi frequency of blue-sideband transition) and $\Delta(t)$ (laser detuning from a motional sideband) are controlled as the red and blue curves. The phase $i \beta$ in $\Omega(t)$ is the counter-diabatic term to suppress the transition during the evolution. Here $\Omega_{0} = (2 \pi) 38.5$ kHz, $\beta = 0.075$, and $\Delta_{0} = 1.6 \Omega_{0}$. (b) Experimental demonstration of the uniform blue-sideband transitions. The total time to drive the transition is $91 \mu$s for any $\ket{n}$ up to $n=5$, which is about 7 times the $\pi$-pulse for an $n=0$ blue-sideband transition. Adapted from Ref.~\cite{Um16Phonon}.
}
\end{figure}

The $\hat{S}^+$ and $\hat{S}^-$ operations can be implemented using the uniform blue-sideband transition. The typical blue sideband transition in Eq. (\ref{eq:RsbBsb}) drives the transition $\ket{\downarrow, n} \leftrightarrow \ket{\uparrow, n+1}$ with an oscillation frequency of $\sqrt{n+1} \eta \Omega$, due to the $n$-dependence of the $\hat{a}^{\dagger}$ and $\hat{a}$ operators. However, a scheme involving an adiabatic passage with a counter-diabatic term, while compensating for AC-stark shift and dynamical phase, can efficiently implement a uniform blue-sideband transition that is independent of $n$. In the scheme, the amplitude and detuning of driving the blue-sideband transition are modulated as $\Omega (t) = \Omega_{0} \left[\sin(\pi t/T) + i\beta \right]$ and $\Delta(t)=\Delta_{0} \cos(\pi t/T)$, respectively, where $i\beta$ is an additional counter-diabatic term that is necessary for keeping the state of the system in the instantaneous ground state during the fast driving \cite{Xi2010Shortcut,Bason2012High,Zhang2014Realization,an2016Shortcuts}. On top of these controls, the sign of $\Omega(t)$ and the control of $\Delta(t)$ are inverted and reversed in the middle of the sequence as shown in Fig. \ref{fig:UniformBSB}(a) in order to offset the effects of the AC-stark shift and dynamical phase.

The total duration of a uniform blue-sideband transition for a complete population-transfer is about 7 times longer than the $\pi$-pulse of the blue-sideband transition on $\ket{n=0}$, which is comparatively faster and more efficient than the standard adiabatic scheme~\cite{Zhang2014Realization} and rapid adiabatic passages~\cite{Wunderlich2007Robust}. 
Fig.~\ref{fig:UniformBSB}(b) shows the experimental performance of the uniform transfer of population from $\ket{\downarrow, n}$ to $\ket{\uparrow, n+1}$ in the range from $n=0$ to 5. 

\begin{figure}
\centering
\includegraphics[width=\columnwidth]{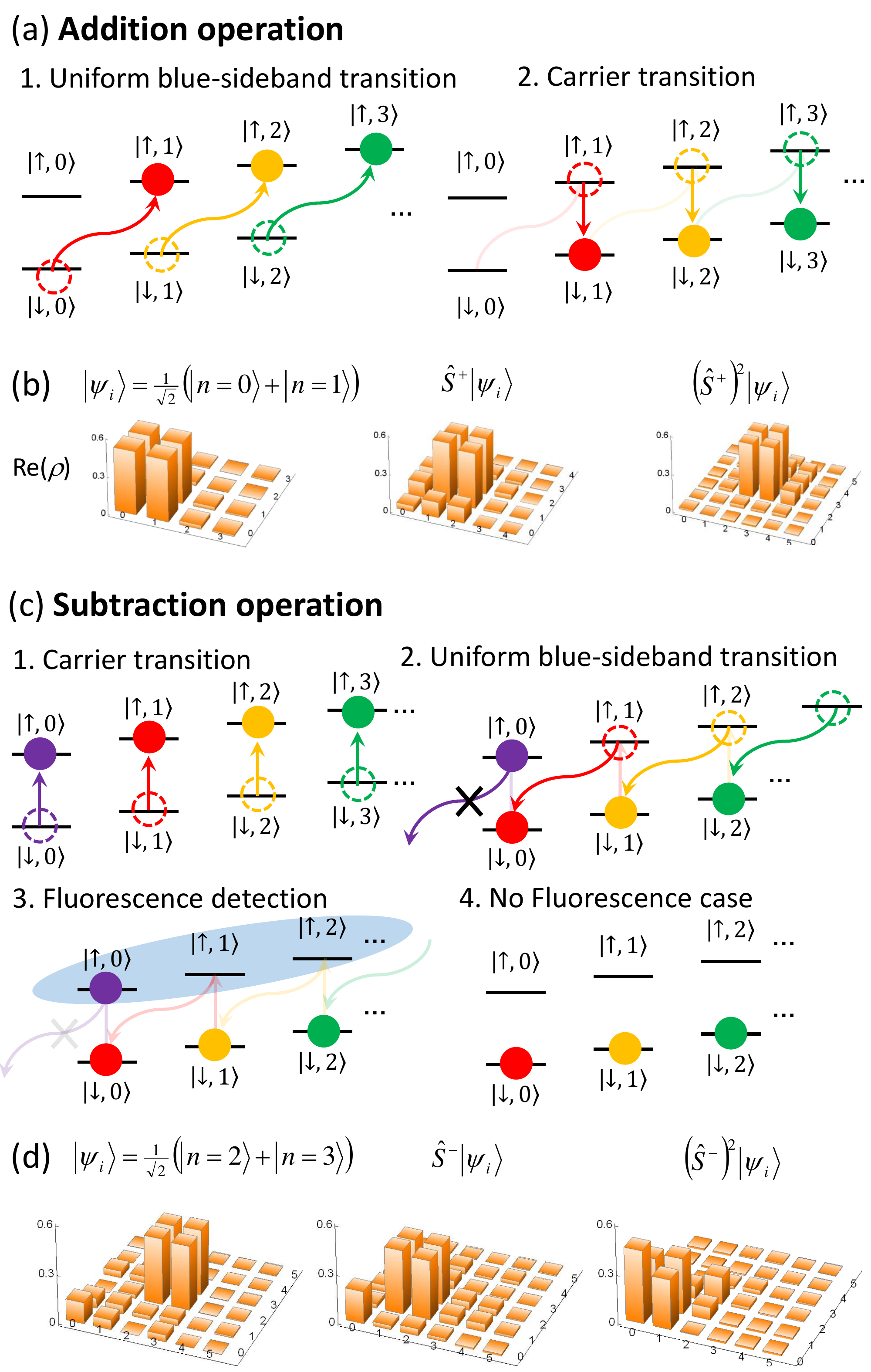}
\caption{
\label{fig:AddSubtract}
(a) Sequence of an addition operation, composed of a uniform blue-sideband transition followed by a $\pi$-pulse driving the carrier transition. (b) Additions on a superposition state $\ket{\psi_{i}}=\frac{1}{\sqrt{2}}\left( \ket{n=0}+\ket{n=1}\right)$. The reconstructed density matrices indicate a fidelity of 0.99 (0.01) for the initially prepared state, and a corresponding fidelity of 0.96(0.01) and 0.92(0.01) for one and two additions, respectively. (c) Sequence of a subtraction operation, realized by the reverse of addition and fluorescence detection. (d) Subtraction on a superposition state $\ket{\psi_{i}}=\frac{1}{\sqrt{2}}\left( \ket{n=2}+\ket{n=3}\right)$. The reconstructed density matrices indicate a fidelity of 0.96 (0.01) for the initially prepared state, and a corresponding fidelity of 0.77(0.02) and 0.83(0.01) after one and two subtractions, respectively. Adapted from Ref.~\cite{Um16Phonon}.
}
\end{figure}

The addition operation $\hat{S}^{+}$ is realized by the sequence shown in Fig. \ref{fig:AddSubtract}(a). The subtraction operation $\hat{S}^{-}$ is realized by reversing the sequence of the addition operation, followed by fluorescent detection as shown in Fig. \ref{fig:AddSubtract}(c). The zero phonon state $\ket{n=0}$ is eliminated after the subtraction, which is implemented by the conditional measurement in the experimental scheme. After the detection sequence, only data with no fluorescence observed are collected. The success rate is given by the probability of the non-zero phonon states. 

The addition and the subtraction scheme deterministically adds and removes one quanta independent of the initial phonon number state while maintaining quantum coherence. Various initial states including a superposition of two Fock states and coherent states have been used to test the performance of the addition and the subtraction operations~\cite{Um16Phonon}. For the example of Fock-state superposition shown in Fig. \ref{fig:AddSubtract}(b,d), coherences represented by the off-diagonal terms of the density matrix clearly remain after addition and subtraction operations. The density matrix is constructed by using the iterative maximum-likelihood algorithm~\cite{Jezek2001Iterative} after displacing the state by 8 different angles with an amplitude of $\alpha \sim 0.8$, followed by determining the phonon number distributions (Eq.~\ref{eq:bsbfit})~\cite{leibfried1996experimental}.

We note that the arithmetic operations discussed are non-Gaussian operations. However, it is not clear whether the nonlinearity introduced by them are strong enough to implement universal quantum computation and quantum simulation with phonons. These operations can make up a part of the phonon number resolving detection as discussed in section III, which enables phononic systems to perform algorithms of the boson-sampling type, without facing limitations of detection efficiency.

\subsection{Nonlinear coupling between harmonic oscillators} 

So far in this review we have assumed that the modes of motion in a system of trapped ions are described by a simple model of a quantum harmonic oscillator, and these modes do not interact with each other in the absence of external forces. This would indeed be the case if the interaction potential between the ions were harmonic, and the force acting on the ion was proportional to the ion displacement from equilibrium~\cite{Landau1976Mechanics}. While these assumptions are true for the trapping potential in a large ion trap where the potential changes smoothly on a time scale comparable to the amplitude of the ion motion, the Coulomb interaction between ions with inter-ion separation of a few microns introduce noticeable anharmonicity, even when the amplitude of the ion's motion is comparable to the size of ion wave packet in the ground state of motion $\sim \sqrt{\hbar / M \omega_m} $, typically on the order of $\sim 10$\,nm. This leads to nonlinear coupling between the normal modes for the oscillator energies on the order of single quanta. 

Anharmonicity gives rise to effects which are studied in the field of nonlinear quantum optics, and are described by the well known models of degenerate and non-degenerate parametric amplifier, as well as the cross Kerr effect. However, unlike the case in optical domain, this nonlinear coupling is strong and deterministic, and manifests itself at the level of a single quanta. In this section we show how all of this interactions can be observed for the system of trapped ions.

\subsubsection{Degenerate parametric interaction}
\begin{figure}[tb]
\centering
\includegraphics[width=\columnwidth]{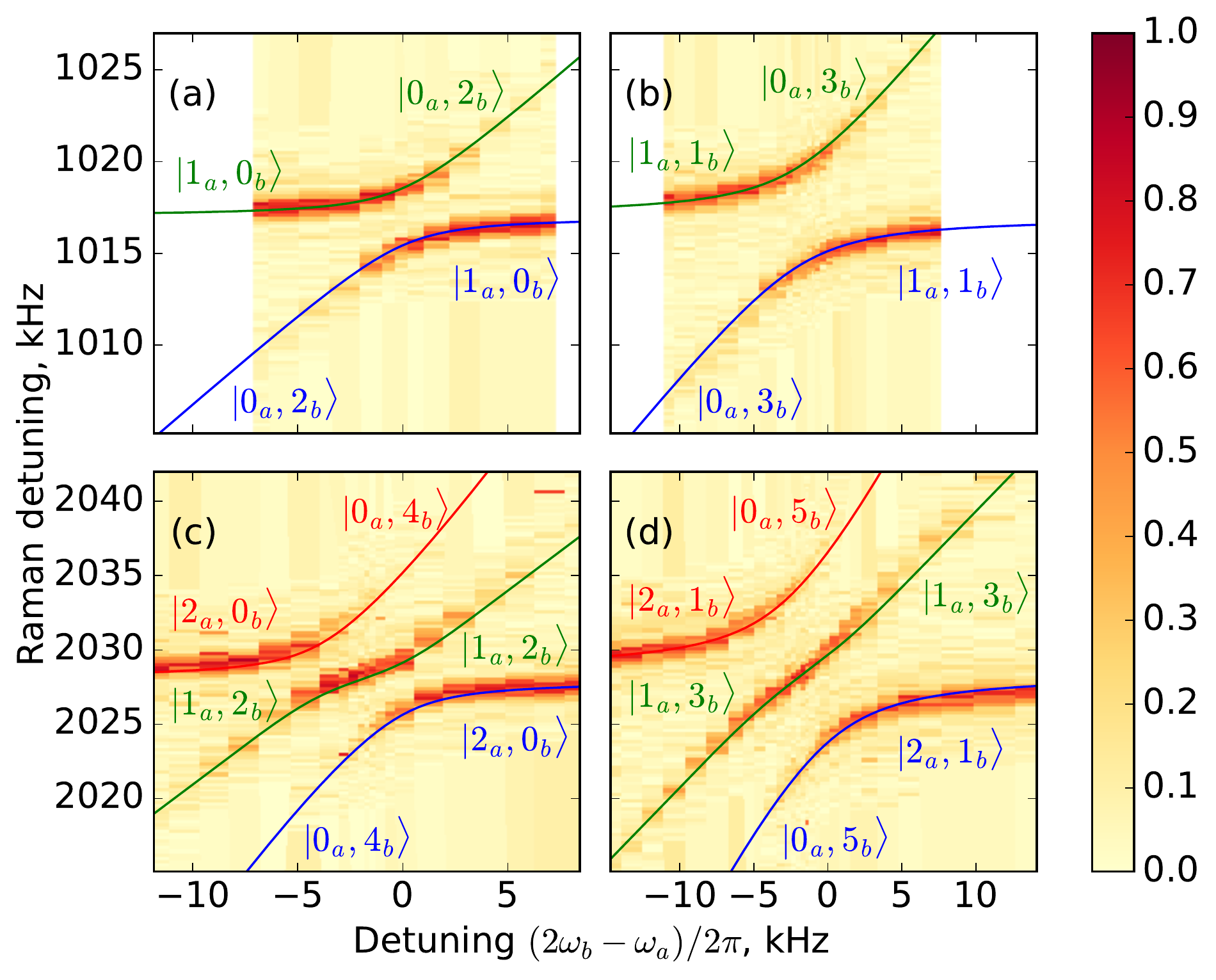}
\caption{
\label{fig:anticross}
Observation of avoided level crossing between various motional eigenstates as a function of the degenerate parametric interaction detuning from resonance $\delta$. Shown are the probability to detect the ion in the $\ket{\uparrow}$ state as a function of detuning $\delta_{\rm degen}$ and Raman beams. Detuning of Raman beams are scanned around the (a,b) first order and (c,d) second order blue sideband transition of the axial mode, for an initially prepared (a) $\ket{1_a,0_b}$, (b) $\ket{1_a,1_b}$, (c) $\ket{2_a,0_b}$, and (d) $\ket{2_a,1_b}$ state.
Figure adapted from Ref.~\cite{Ding2017Cross}.
}
\end{figure}

To estimate the strength of the nonlinear coupling, consider a simple case of two co-trapped ions having the same mass $M$ and charge $e$ in a linear Paul trap, with single-ion secular frequencies $\omega_{\rm{X}}$, $\omega_{\rm{Y}}$, $\omega_{\rm{Z}}$. The potential energy of the system is~\cite{Marquet2003phonon,Nie2009Theory} 
\begin{eqnarray} 
\label{eq:energy}
 V &=& M \omega_{\rm{X}}^2 (X^2+x^2) + M \omega_{\rm{Y}}^2 (Y^2 +y^2) + M \omega_{\rm{Z}}^2 (Z^2+ z^2)\nonumber  \\ 
 & & + \frac{e^2}{8 \pi \epsilon_0} \frac{1}{\sqrt{x^2 + y^2 + z^2}},
\end{eqnarray}
where $\epsilon_0$ is the permittivity of free space, $X, Y, Z$ are the center-of-mass coordinates, and $x, y, z$ are half the separation between the ions along the direction of principle trap axes. When $\omega_{\rm{Z}} < (\omega_{\rm{X}}, \omega_{\rm{Y}}$), the ions crystallize along the axial ($z$) direction at an equilibrium distance  $z_0$  from the trap center. According to equation~(\ref{eq:energy}), the motion of the center-of-mass modes is harmonic,  but the out-of-phase modes are coupled by the  Coulomb interaction. For small axial displacement $u=z-z_0$ and keeping only terms up to the  third order that contribute to the coupling between the $x$ and $z$ modes, the potential energy becomes~\cite{Nie2009Theory}

\begin{equation}\label{eq:pot}
V = M\omega_a^2 u^2 + M\omega_{b}^2 x^2 +\frac{M\omega_a^2}{z_0} x^2 u + ... \, \nonumber.
\end{equation}


Here $\omega_a=\sqrt{3}\omega_{\rm{Z}}$ and $\omega_b=\sqrt{\omega_{\rm{X}}^2-\omega_{\rm{Z}}^2}$ are the axial ``stretch" and radial ``rocking"  mode frequencies for the out-of-phase motion. If the trap frequencies satisfy the condition $\delta_{\rm degen} = \omega_a - 2 \omega_b \simeq 0$, we can apply the standard transformations  $\hat{u}=(\hbar/4M\omega_a)^{1/2}(\hat{a}+\hat{a}^\dagger)$ , $\hat{x}=(\hbar/4M\omega_b)^{1/2}(\hat{b}+\hat{b}^\dagger)$ and express the Hamiltonian in the rotating wave approximation as 
\begin{equation}
\label{eq:ham}
\hat{H}=\hbar\omega_a \hat{a}^{\dagger}\hat{a} + \hbar\omega_b \hat{b}^{\dagger}\hat{b}+\hbar\xi_{\rm d}\left(\hat{a}\hat{b}^{\dagger\,2} + \hat{a}^{\dagger} \hat{b}^2 \right),
\end{equation}
where $\hat{a}^{\dagger},\hat{a},(\hat{b}^{\dagger},\hat{b})$ are the phonon creation and annihilation operators in axial (radial) mode. The first two terms in equation~(\ref{eq:ham}) describe harmonic motion in the axial (radial) mode with the frequency $\omega_a\,(\omega_b)$ and the third term couples these modes with the coupling coefficient given by 
\begin{equation}
\label{eq:coupling}
\xi_{\rm d}= \frac{1}{8 z_0}\sqrt{\frac{\hbar \omega_a^3}{M\,\omega_b^2}} \, .
\end{equation}
The coupling is nonlinear: if the resonance condition $\delta_{\rm degen}=0$ is met, one axial phonon is converted into a pair of radial phonons and vice versa. It is described by a Hamiltonian identical to the degenerate parametric down conversion Hamiltonian in quantum optics.
Similar coupling also exist between modes of motions of three trapped ions.

When $\xi_{\rm d}\neq0$, the ``bare'' energy eigenstates of the axial and radial modes are mixed such that eigenstates of Hamiltonian Eq.~(\ref{eq:ham}) have non-zero projections along both modes. This coupling manifests itself in the form of an avoided crossing~\cite{Ding2017Quantum,Ding2017Cross} as $\delta_{\rm degen}\rightarrow0$, shown in Fig.~\ref{fig:anticross}. 

\subsubsection{Cross Kerr effect}
If the resonance conditions are not met and energy exchange between the modes is not possible, this coupling manifests itself as a shift in the frequency of a motional mode, which is proportional to the number of phonons in the other motional mode and can be described by an effective cross-Kerr interaction~\cite{Nie2009Theory}. A small shift (about 20 Hz/phonon) of this origin was first observed in \cite{Roos2008Nonlinear} using a Ramsey type experiment. By tuning the mode frequencies close to the parametric resonance, an order of magnitude increase in sensitivity to phonon number ($\sim300$ Hz/phonon) was observed in~\cite{Ding2017Cross}.  

When the axial and radial modes are far detuned from each other, i.e., $\delta_{\rm degen} \gg \xi_{\rm d}$,  the coupling manifests itself as frequency shifts of the modes.  In the lowest non-vanishing order of perturbation theory, shift of the axial mode $\delta E = \delta E_{n_a+1,n_b} - \delta E_{n_a, n_b} = -2 (2n_b+1) \xi^2/\delta_{\rm degen}$ is proportional to the number of phonons $n_b$ in the radial mode~\cite{Nie2009Theory}.  Here $\delta E_{n_a,n_b} = \sum_{k_a, k_b} |\langle n_a, n_b | H | k_a, k_b \rangle|^2 /( E_{ n_a, n_b} - E_{ k_a, k_b} )$, and $E_{ j_a, j_b} = j_a  \omega_{a} + j_b \omega_{b}$. This shift deviates from a linear relation due to finite value of $\delta_{\rm degen}$.  However, exact diagonalization of the Hamiltonian Eq. (\ref{eq:ham}) confirms that the frequency shift is still a monotonous function of the radial phonon number $n_b$~\cite{Ding2017Cross}. 

\begin{figure}[tb]
\centering
\includegraphics[width=\columnwidth]{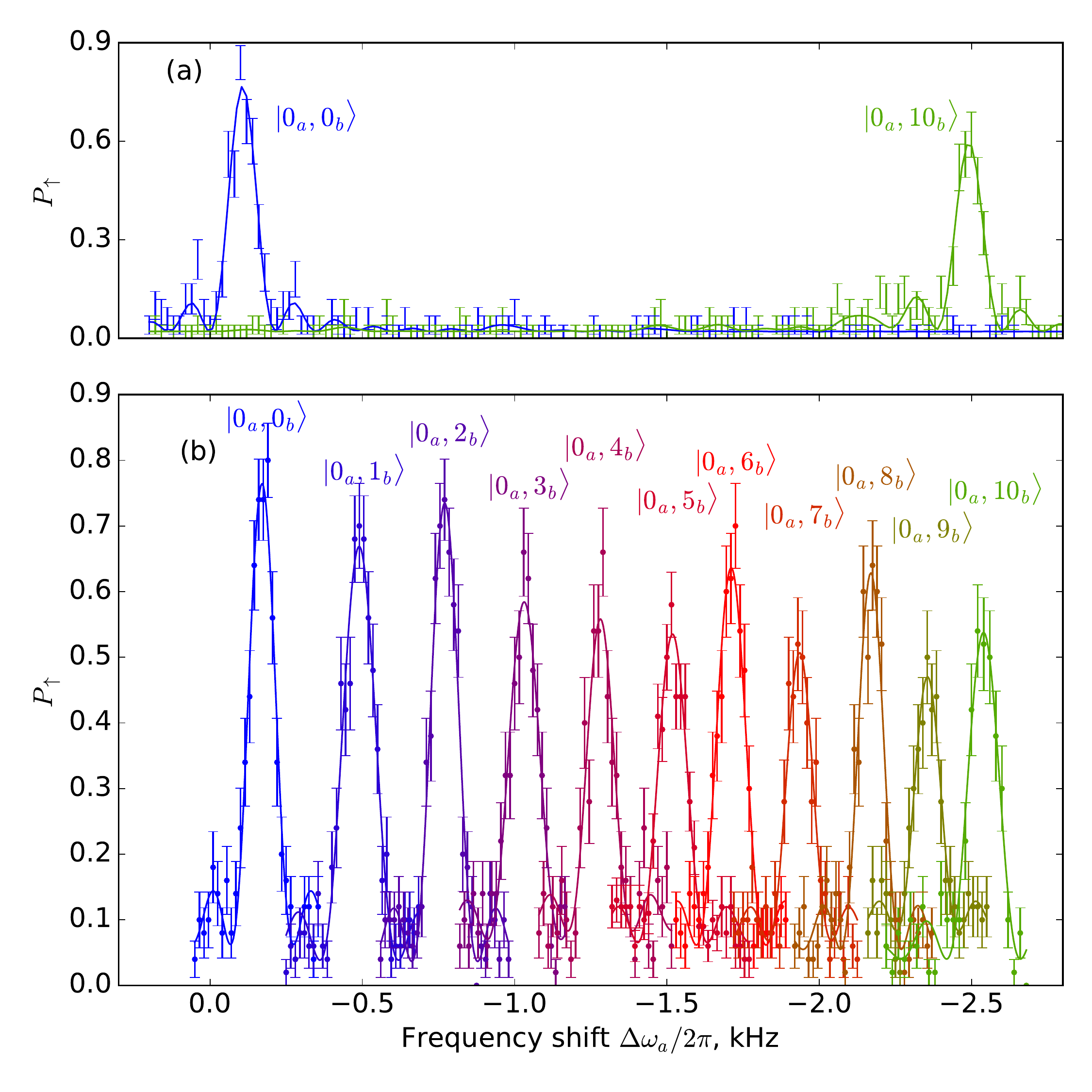}
\caption{\label{fig:fock_peaks}
(a) Full frequency scan of the axial blue sideband when the system is prepared 
in Fock states $|0_a, 0_b\rangle$ (blue) and $|0_a, 10_b\rangle$ (green).
For the state $|0_a, 10_b\rangle$, the reduced height of the main peak and residual 
population detected in the states with $n_b<10$ are due to imperfections of the state preparation
that, according to the fit, populates the state $|0_a, 10_b\rangle$ with probability 
$p_{10} = 0.80(2)$, and 
leaves significant populations $p_{9} = 0.06(2)$, and  $p_{8} = 0.06(2)$ in the states
$|0_a, 9_b\rangle$ and $|0_a, 8_b\rangle$. Preparation of state $|0_a, 0_b\rangle$
populates other Fock states with the probability $<0.04$. 
(b) Measured position of the axial motional sideband for
the Fock states $|0_b\rangle$ to $|10_b\rangle$ prepared in the radial mode. The axial mode is 
initially prepared in the vacuum state $|0_a\rangle$.
Only part of the data around the largest peak is shown for each state~\cite{Ding2017Cross}.
}
\end{figure}

To observe this effect, three $^{171}$Yb$^+$ ions are trapped in a standard linear rf-Paul trap~\cite{Ding2017Quantum,Ding2014Microwave} with single ion secular frequencies $(\omega_{\rm{X}}, \omega_{\rm{Y}}, \omega_{\rm{Z}}) = 2\pi \times (1042, 979, 587)$ kHz. The axial trapping frequency $\omega_{\rm{Z}}$ is fixed and the radial trapping frequencies $\omega_{\rm{X}}, \omega_{\rm{Y}}$ are fine tuned by adjusting offset voltages applied to the trap electrodes. For convenience, two of the ions are optically pumped to a metastable long-lived $^2F_{7/2}$ state and remain dark throughout the experiment. Three Raman beams in total are used to address all nine motional modes along all three principal axes of the trap for sideband cooling, motional state preparation, and detection. To ensure that all the motional modes can be addressed by the Raman lasers, the bright ion is always positioned at the edge of the crystal. 

As shown in Fig. \ref{fig:fock_peaks}, the observed frequency shifts are on the order of 300 Hz / phonon and are in good agreement with theoretical predictions calculated by diagonalization of the Hamiltonian Eq. (\ref{eq:ham}). The peaks are clearly separated from each other, which makes it possible to efficiently detect the phonon number distribution for the radial mode. 

\subsubsection{Nondegenerate parametric interaction}
Nonlinear coupling between the modes can also lead to nondegenerate parametric interaction between mechanical modes of motion in the system of trapped ions. Motion of three ions in the approximately harmonic trap is decomposed into a set of decoupled normal modes: the motion along each trap axis is described by the center-of-mass mode with the eigenvector $e_{cm} = (1, 1, 1) / \sqrt{3}$, the ``tilt'' mode $e_{t} = (-1, 0, 1) / \sqrt{2}$, and the ``zigzag'' mode $e_{zz} = (-1, 2, -1) / \sqrt{6}$. 

Symmetry and energy conservation restricts the set of modes that are coupled. The center-of-mass modes cannot be coupled to any other mode, since they are purely determined by the trap potential and do not depend on the Coulomb interaction between the ions. The energy conservation condition is fulfilled in particular by tuning the mode frequencies such that they satisfy
\begin{equation}
\label{eq:nondegen_resonancecondition}
\omega_a = \omega_b + \omega_c,
\end{equation}
where $\omega_b\neq\omega_c$.
When $\omega_{\rm{Z}} = 0.556\,\omega_{\rm{X}}$, this condition is satisfied for the axial zigzag mode with the frequency  $\omega_a = \sqrt{29/5}\,\omega_{\rm{Z}}$, and the tilt and zigzag modes along the $x$-radial direction with the frequencies $\omega_{b} = \sqrt{\omega_{\rm{X}}^2 - \omega_{\rm{Z}}^2}$ and $\omega_c = \sqrt{\omega_{\rm{X}}^2 - 12\,\omega_{\rm{Z}}^2 / 5}$, respectively. Detuning from resonance Eq.~(\ref{eq:nondegen_resonancecondition}) is defined as $\delta_{\rm nondegen} = \omega_a - \omega_b - \omega_c$.

\begin{figure}[tb]
\centering
\includegraphics[width=0.8\columnwidth]{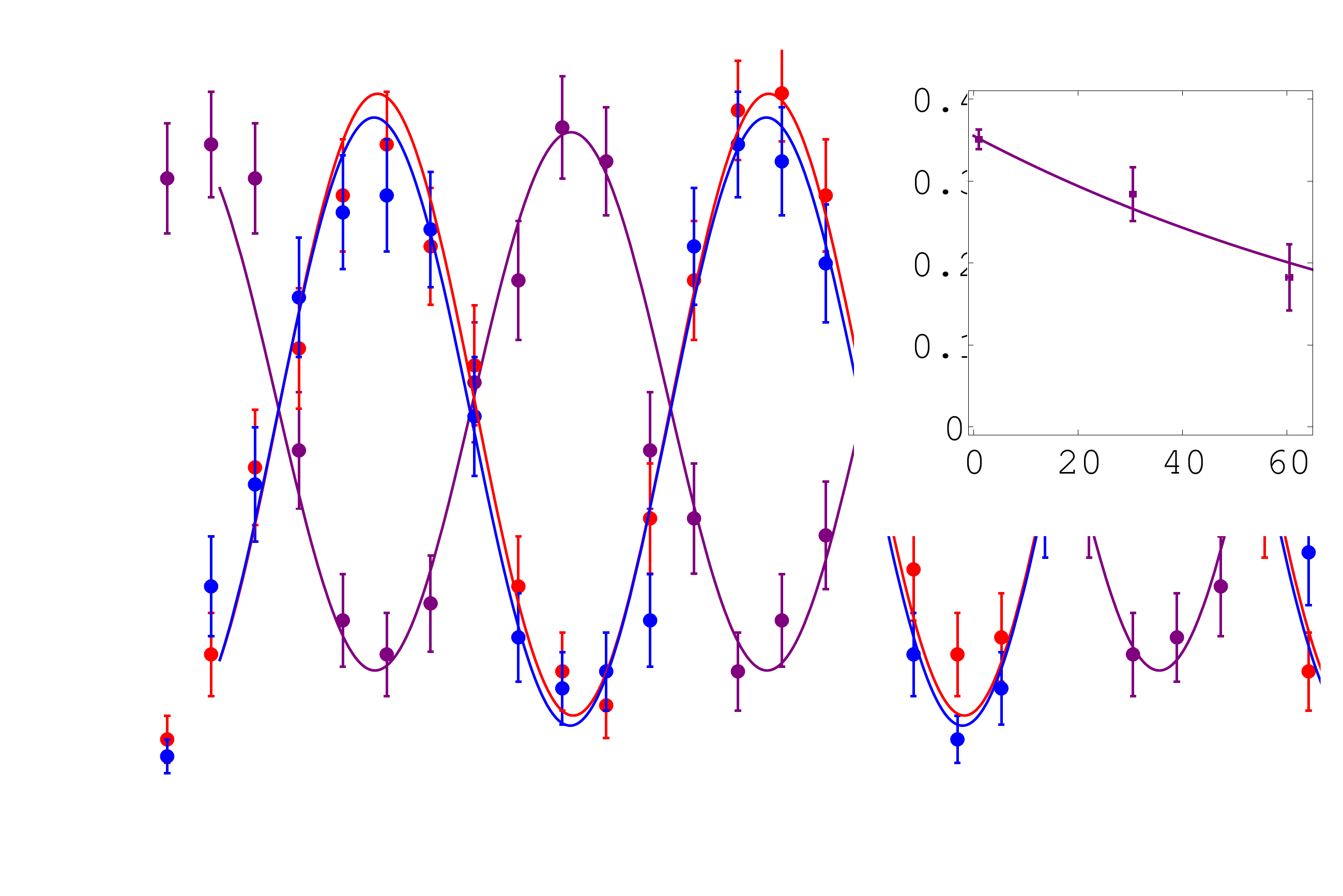}
\caption{\label{fig:trilinear_coupling}
Coherent energy exchange between three modes of motion coupled by the non-degenerate parametric interaction. The purple, red and blue dots represent the probabilities to drive the corresponding red sideband transitions of the axial zig-zag ($\omega_a=1414$ kHz), the radial tilt ($\omega_b=878$ kHz) and the radial zig-zag ($\omega_c=536$ kHz) modes, respectively, as functions of three-mode interaction time $\tau$. The purple, red and blue lines are the corresponding fits. The oscillations decay with a time constant of $0.11(2)$ s, which corresponds to more than 300 cycles of energy exchange~\cite{Ding2018Trilinear} }
\end{figure}

A signature of the cross-mode coupling is a coherent energy exchange between different modes under the resonance condition~\cite{Ding2018Trilinear}. We start with one phonon in the axial mode and zero phonon in both radial modes, and then adjust the detuning $\delta_{\rm nondegen}=0$ (resonance). After a wait duration $\tau$, we tune the detuning back to its initial value $\delta_{\rm nondegen}=-2 \pi \times 44~\text{kHz}$ and measure the phonon populations in all the three modes. We observe a coherent energy exchange between the axial and the two radial modes with a frequency $2.801(2)~\text{kHz}$ (Fig.~\ref{fig:trilinear_coupling}). The amplitude of oscillations reduces to $1/e$ on a time scale of 0.11(2) s, which corresponds to about 330 oscillation cycles. 

\section{VI. Examples of applications}

\subsection{Demonstration of the GKP code} 

Quantum error correction~\cite{shor1995scheme, steane1996multiple} is essential for the realization of scalable universal quantum computation~\cite{nielsen2002quantum}. The essence of various error-correction codes is to encode a logical qubit with a larger Hilbert space so as to protect quantum information by redundancy. Instead of using multi-qubit systems, Gottesman, Kitaev and Preskill (GKP)~\cite{gottesman2001encoding} proposed an error-correction code that encodes a logical qubit in the Hilbert space of a single harmonic oscillator. In this error-correction code, the logical code words are coherent superpositions of an infinite series of infinitely squeezed states, i.e. eigenstates of the position or momentum operators, and their Wigner functions form an infinite lattice in the phase-space of the oscillator. 

Recently, the encoding, quantum control, and readout of a GKP logical qubit has been demonstrated using a trapped \Ca ion~\cite{fluhmann2019encoding}. The logical qubit is encoded in a vibrational mode that is manipulated by laser-induced coupling with the internal states of the trapped ion. To ensure normalization and minimal effects from motional dephasing, the logical computational basis states are constructed from a finite series of squeezed states. Specifically, the approximate state to the logical GKP $\ket{0}_L$ state is
\begin{eqnarray}
\ket{0}_L=\sum_{k=-K}^Kc_k\hat {\mathcal D}\left(kl\right)\ket{r},
\end{eqnarray}
where $\hat{\mathcal D}\left(\cdot\right)$ and $\ket{r}$ are the displacement operator and the position-squeezed state, respectively. The exact logical GKP  state is recovered when $r$ and $K$ both go to infinity. The pauli operators $\hat X_L$, $\hat Y_L$ and $\hat Z_L$ in the logical space are displacement operators, i.e. $\hat X_L=\hat{\mathcal D}\left(\frac{l}{2}\right)$, $\hat Y_L=\hat{\mathcal D}\left(-\frac{l}{2}-\frac{i\pi}{l}\right)$, and $\hat Z_L=\hat {\mathcal D}\left(\frac{i\pi}{l}\right)$.

The preparation of $\ket{0}_L$ starts with obtaining $\ket{r}$ by reservoir engineering~\cite{kienzler2015quantum}, with $r\simeq0.9$. The superposition of displacement operation is then realized by modular variable measurement (MVM)~\cite{fluhmann2018sequential}. Two essential ingredients of the MVM sequence are the laser-induced state-dependent force, and fluorescence detection of the internal states. Repeating the MVM sequence twice and conditioned on detecting no fluorenscence, the authors prepare the three-component $\ket{0}_L$ and the four-component $\ket{1}_L$ states and measured the marginal position and momentum probability distributions. The average logical state preparation and readout fidelity is reported to be $87.3\%$.

The authors also demonstrate the implementation of arbitrary control and measurement of logical operators. First of all, the displacement operators, i.e. the logical Pauli operators, are implemented by applying a resonant voltage signal to one of the trapping electrodes~\cite{leibfried1996experimental}. Arbitrary control is realized by the gate teleportation technique~\cite{knill1998resilient}, which is widely used in error-correction codes. Here in the trapped-ion context, the gate teleportation is implemented by performing a single-qubit rotation on the internal states of the trapped ion before the MVM sequence. The measurement of logical operators, including Pauli operators and stabilizers, can be mapped to the readout of the MVM sequence by performing appropriate logical operations. The authors characterized the performance of the logical operations by quantum process tomography. The process fidelities for the logical Pauli operators are $97\%$ and those for partial rotations are $90\%$.


\subsection{NOON state}
A maximally entangled state of bosonic particles can be in a form of a NOON state. The NOON state of identical bosons~\cite{Pezze2009Entanglement,Giovannetti2011Advances} provide the ultimate Heisenberg limit, which, for $N$ bosons, has the form \cite{Sanders89Quantum} 
\begin{equation}
\Ket{\psi_{{\rm NOON}}}=\frac{1}{\sqrt{2}}\left(\Ket{N,0}+e^{iN\varphi_{\rm S}}\Ket{0,N}\right),\label{eq:NOON}
\end{equation}
where the relative phase $\varphi_{\rm S}$ between the two terms of the state is linearly proportional to $N$ in an interferometer, showing the Heisenberg scaling for parameter estimation. For photonic systems, experiments have demonstrated NOON states with particle numbers up to $N=5$ \cite{Kok2002Creation,Mitchell2004Super,Walther2004De,Nagata2007Beating,Dowling2008Quantum,Afek2010High,Wolfgramm2013Entanglement,Liu2015Demonstration}. For distinguishable particles, up to $10$ photons and $14$ ions have been prepared in the closely-related GHZ\ states \cite{wang2016experimental,Monz2014Qubit}. NOON states have also been demonstrated in nuclear spins (NMR) \cite{Jones2009Magnetic}, atomic spin waves \cite{Chen2010Heralded}, and microwave photons in superconducting systems \cite{Wang2011Deterministic}. 

\begin{figure}[tb]
\centering
\includegraphics[width=\columnwidth]{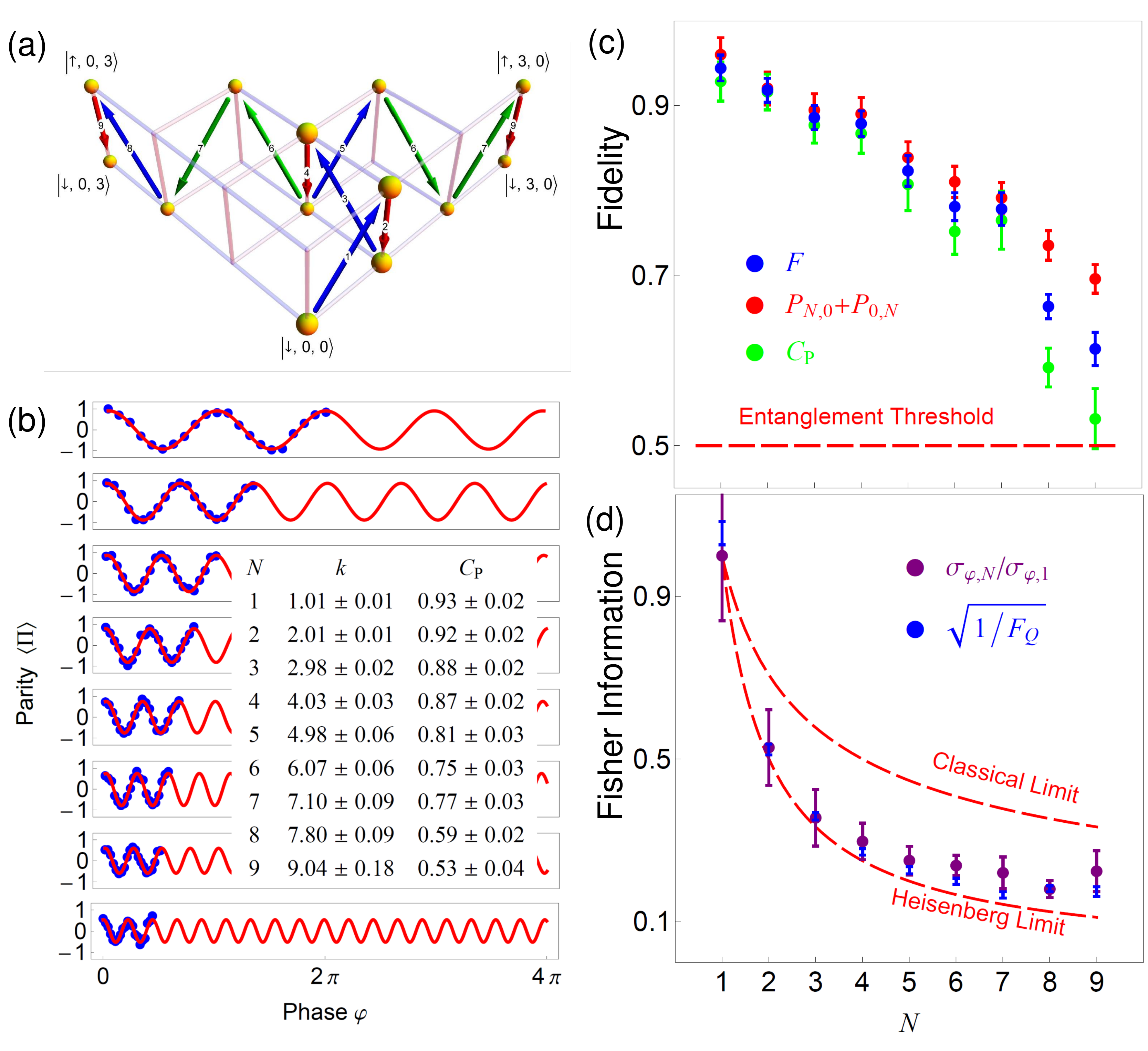}
\caption{\label{fig:NOON}
(a) Generation sequence of the NOON state of $N=3$. The blue arrows indicate blue-sideband transitions, the red arrows indicate carrier transitions and the green arrows indicate composite-pulse operations. The numbers on the arrows denote the sequence of the operations. (b) Parity oscillations of the generated NOON states from $N=1$ to $N=9$. The blue dots are experimental data, and the red lines are fitting curves with $\left\langle \Pi\left(\varphi\right)\right\rangle =A\cos k\varphi+B\sin k\varphi+C$, and $C_{\mathrm{P}}=\sqrt{A^{2}+B^{2}}$. (c) The experimental results of fidelity as well as $C_{\mathrm{P}}$ and $P_{N,0}+P_{0,N}$. The error bars of $C_{\mathrm{P}}$ are derived from fitting error and those of $P_{N,0}+P_{0,N}$ from shot-noise error. (d) The quantum Fisher information of the generated states. Figure adapted from Ref.~\cite{Zhang2018Experimental}}
\end{figure}

A deterministic scheme to generate phononic NOON states with arbitrary number of bosons $N$ \cite{Zhang2018Experimental} has been developed based on blue-sideband transitions of Eq. (\ref{eq:RsbBsb}), where two radial modes of an \Yb ion trapped in a 3D harmonic oscillator has been used, but the generation scheme can be similarly applied to different atomic ions with more modes. In Fig.~\ref{fig:NOON}(a), the generation scheme is illustrated for the case of $N=3$ NOON state, $\Ket{3,0}+\Ket{0,3}$, as an example. With the pulse sequence, the NOON state up to $N=9$ has been generated, which is mainly limited by experimental imperfections \cite{Zhang2018Experimental}.  

The sensitivity of the phase estimation with the NOON states increases as the number of phonons $N$ increases. The phase between X and Y modes can be measured by the interference through the beam splitting operation of Eq. (\ref{eq:BS}). In the experiment, the beam splitting operation is realized by linear combinations of input modes \cite{Zhang2018Experimental}. In the experiment, the parity of phonons in the output modes of the generated state is measured depending on the value of $\varphi$, which provide the phase sensitivity proportional to $N$ as 
\begin{equation}
\left\langle \Pi\left(\varphi\right)\right\rangle =C_{\mathrm{P}}\cos N\varphi.\label{eq:parity}
\end{equation}
Fig. \ref{fig:NOON}(b) shows the experimental results of the parity oscillations from $N=1$ to $N=9$ of the generated NOON states. As shown in the fitting parameter $k$, the enhancement of the phase sensitivity is in agreement with $N$ within 2.6\% deviation. As $N$ increases, the contrast $C_{{\rm P}}$ decreases due to experimental imperfections. However, it is shown that even for $N=9$, the contrast is over 0.5, which indicates the existence of quantum entanglement in the state.

The fidelity $F$ and the quantum Fisher information of the generated NOON state are experimentally measured and studied \cite{Zhang2018Experimental}. As shown in Fig. \ref{fig:NOON}(c), the fidelities of the NOON states up to $N=9$ are larger than 0.5, which confirms genuine multi-party entanglements of the prepared states. The Heisenberg scaling of the lower bound of the sensitivity in the phase estimation is studied through the quantum Fisher information, which provides the best possible precision on a parameter estimation given by $1/\sqrt{F_{Q}}$ \cite{Braunstein1994Statistical,Cooper2011Towards}, known as the Cram\'{e}r-Rao bound. For $N$ particles without entanglement, the best possible measurement scales as $1/\sqrt{N}$ and for the NOON state, the lower bound of the precision scales as $1/N$, the Heisenberg limit. As shown in Fig. \ref{fig:NOON}(d), the lower bound of the phase uncertainty, $1/\sqrt{F_{Q}}$, of the generated states from $N=2$ to $N=9$ violate the classical bound and approaches the Heisenberg limit. 

\subsection{Quantum Thermodynamics}
\subsubsection{Test of quantum Jarzynski equality} 
There is increasing interest in non-equilibrium dynamics at the microscopic scale, spanning across various aspects of quantum physics, thermodynamics, and information theory as technologies of the experimental control at such scales have rapidly been developing. Most principles in non-equilibrium process are represented in the form of an inequality. In the classical regime, a remarkable equality \cite{Jarzynski1997Nonequilibrium} that relates the free-energy difference to the exponential average of the work done on the system was found: 
\begin{eqnarray}
\avg{e^{-(W-\Delta F)/k_{\rm B}T}}=1,  
\label{eq:Jarzynski}
\end{eqnarray}
where $W$ is the work done on the system, $\Delta F$ is the free-energy difference, $T$ is the initial temperature of the system in thermal equilibrium, and $k_{\rm B}$ is the Boltzmann constant. 
Independent of the protocols in which parameters of a system are varied, the Jarzynski equality Eq.~(\ref{eq:Jarzynski}) always holds, even when the driving is arbitrarily far from equilibrium. This is related in part to the Crooks Fluctuation Theorem~\cite{Crooks1999Entropy}. Experimental tests of the classical Jarzynski equality have been successfully performed in various systems \cite{Hummer2001Free,Liphardt2002Equilibrium,Collin2005Verification,Douarche2005An,Bustamante2005nonequilibrium,Blickle2006Thermodynamics,Harris2007Experimental,Saira2012Test,Jarzynski2011Equalities,Seifert2012Stochastic}.    
In the quantum regime, however, it is difficult to determine the work done in the system due to Heisenberg's uncertainty principle. In an isolated quantum system, the work can be measured by two-point projective measurement over the energy eigenstates \cite{Hanggi2007Fluctuation,Esposito2009Nonequilibrium}, and the classical Jarzynski equality can be extended to the quantum regime \cite{Tasaki2000Jarzynski,Kurchan2000Quantum,Mukamel2003Quantum}. In an harmonic oscillator of a trapped ion system, the projective measurements discussed in section III has been applied to experimentally test the Jarzynski equality in quantum regime.  

The work is performed on the system by shifting the center of the potential by a certain distance in the $X$-direction, which is described by the following time dependent Hamiltonian,    
\begin{eqnarray}
H\left(t\right)&=&\frac{\hat{P}^2}{2 M}+\frac{1}{2} M \omega_{\rm{X}}^2 \hat{X}^2 + f(t) \hat{X},
\label{eq:Ham}
\end{eqnarray}
where $M$ is the mass of the ion, and $\omega_{\rm{X}}$ is the trap frequency along the $X$-axis. The external force $f(t)$, which can be applied by an electric field or by properly adjusted laser beams, shifts the trap center by $-f(t)/M\omega_{\rm{X}}^2$ and reduces the minimum energy by $f^2(t)/2M\omega_{\rm{X}}^2$. When the force $f(t)$ is applied adiabatically, the final state distribution is conserved in the new basis and remains in thermal equilibrium. If the force is instantly applied to the same maximum value, the final states are highly excited, which reveals far-from equilibrium dynamics. The Jarzynski equality in the quantum regime can be verified when the average of the exponentiated work $\avg{\exp{(-W/k_{\rm B}T)}}$, which is measured by $\sum{P\left(W\right) \exp{(-W/k_{\rm B}T)}}$, is independent of any protocol of applying the work including the both cases above.

The essential part of the experimental test in the quantum regime is to perform a two-point projective measurement over the energy eigenstates, which are Fock-states for the harmonic oscillator. 
In detail, the experimental procedure is composed of the following four stages: 1. Preparation of thermal State; 2. Projection to an energy eigenstate; 3. Application of work on the eigenstate; 4. Measurement of final phonon distribution. The procedure is repeated to obtain statistically meaningful results.   

\begin{table}
\caption{Summary for the experimental test of the quantum Jarzynski equality. Here ???}
\label{Tab:Results}
\begin{tabular}{c|c|c|c}
\noalign{\hrule height 0.8pt}
				$\Delta F/(k_{\rm B} T_{\rm eff})$ & \multicolumn{3}{c} {$-\ln\avg{e^{-W_{\rm diss}/k_{\rm B}T_{\rm eff}}}$} \\
\cline{2-4}
									& $\tau=$5 $\mu$s &  $\tau=$25 $\mu$s  & $\tau=$45 \\
\noalign{\hrule height 0.8pt}
-2.63 (316 nK)  &  -0.032($\pm$37) & 0.006($\pm$34)	& 0.042($\pm$52)  \\
-2.13 (390 nK) &  -0.033($\pm$35) & 0.005($\pm$33)	& 0.037($\pm$50) \\
-1.73 (480 nK) &  -0.034($\pm$34) & 0.003($\pm$31)	& 0.031($\pm$48) \\
\noalign{\hrule height 0.8pt}
\end{tabular}
\end{table}

Table \ref{Tab:Results} summarizes the test results of the Jarzynski equality applied to the quantum regime and shows that the Jarzynski estimation gives good prediction of free energy within experimental uncertainties under three different protocols, from three different initial temperatures. In the experiment, the free energy differences $\Delta F$ are -2.63 $k_B T_{\rm eff} ~(T_{\rm eff}=316 \rm nK)$, -2.13 $k_B T_{\rm eff} ~(T_{\rm eff}=390 \rm nK)$, and -1.73 $k_B T_{\rm eff} ~(T_{\rm eff}=480 \rm nK)$ depending on initial temperatures. The experimental data clearly demonstrates the validity of Jarzynski equality when other estimations deviate from the ideal values in far-from equilibrium regime \cite{Jarzynski1997Nonequilibrium}. It is noted that the quantum Jarzynski equality has also been tested in an open quantum system under a decohering heat bath for a spin system of a trapped ion~\cite{smith2018verification}.    

\subsubsection{Quantum absorption refrigerator} 

\begin{figure}[tb]
\centering
\includegraphics[width=\columnwidth]{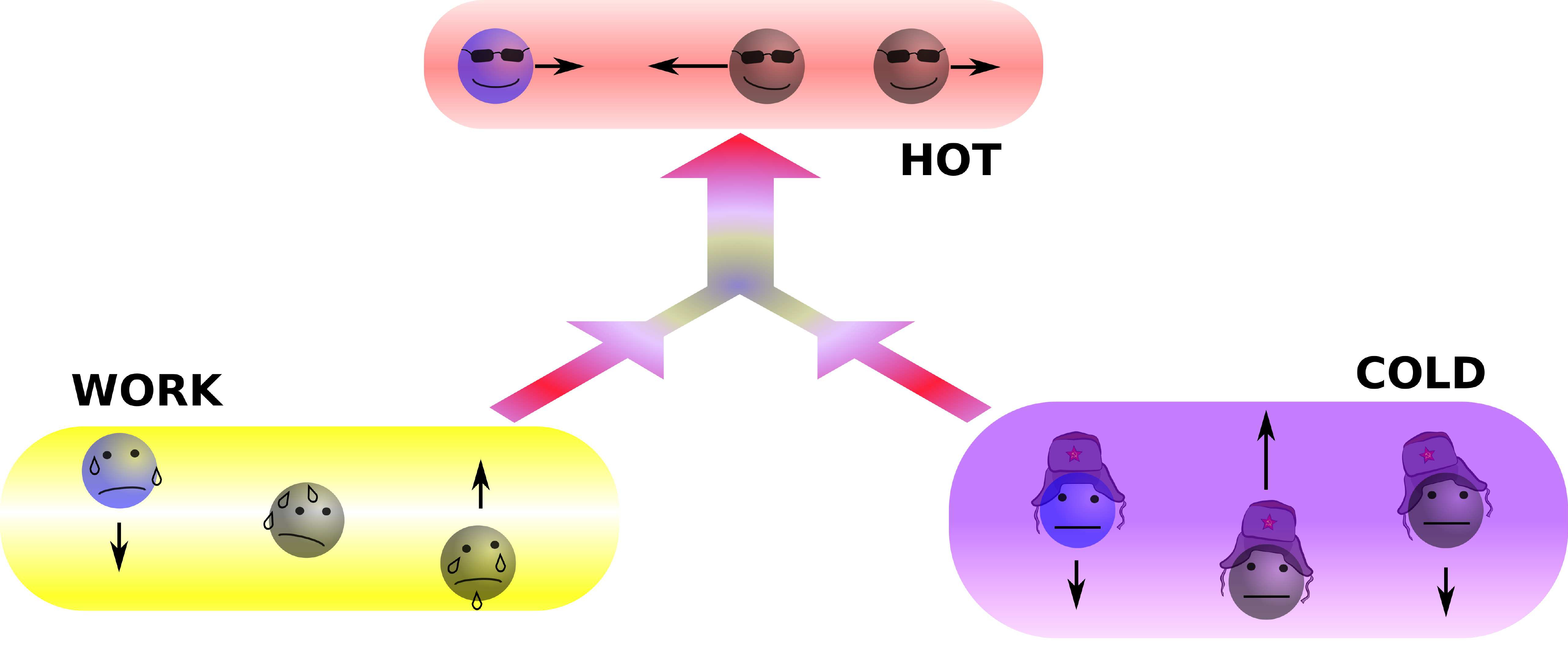}
\caption{
\label{fig:heat_flow}
Illustration of heat flow in an absorption refrigerator. Due to the nature of the interaction among the three bodies (Eq.~\ref{eq:fridge_interaction}), when the work body is at a higher temperature than the hot body, the resulting heat flow causes heat from the cold body to flow to the hot body as well. Thus heating the work body, and hence increasing its temperature, causes the cold body to cool down.
}
\end{figure}

A quantum absorption refrigerator is the simplest example of a quantum heat machine. It consists of 3 heat reservoirs: the hot, cold, and work. They interact in such a way that the heat flow from the work to the hot reservoir cools down the cold counterpart. In the case of trapped ions, collective motional modes of the three-ion crystal serve as the heat reservoirs, and nonlinear coupling among the normal modes of motion are induced by the anharmonicity of Coulomb interactions between the ions. The non-degenerate parametric coupling between the three modes of motion described earlier is sufficient for an experimental realisation of a quantum absorption refrigerator utilizing three modes of motion of trapped Ytterbium ions, one axial and two radials, as the heat bodies (Fig.~\ref{fig:heat_flow}). The radial "zig-zag" (cold) mode represents the heat body that is in contact with a cold environment to be refrigerated, while the axial "zig-zag" (hot) mode of higher frequency represents the heat body in contact with the ambient temperature. The third radial "rocking" (work) mode serves as the heat source that drives the refrigeration of the cold mode, replacing the work reservoir of a conventional refrigerator.

The interaction Hamiltonian in the system of three ions, induced by anharmonicity of the Coulomb repulsion between the ions, has the form~\cite{Marquet2003phonon,Levy2012Refrigerator}
\begin{equation}
\label{eq:fridge_interaction}
\hat{H} =  \hbar \xi_{\rm n} ( \hat{a}_h^{\dagger} \hat{a}_w \hat{a}_c + \hat{a}_h \hat{a}_w^{\dagger} \hat{a}_c^{\dagger} ),
\end{equation}
where the $\hat{a}_i$ ($\hat{a}_i^\dagger$) are the annihilation (creation) operators for the corresponding harmonic oscillators labeled by $i=h,w,c$, and $\xi_{\rm n} = 9 \omega_{\rm{Z}}^2\sqrt{\hbar/M \omega_h \omega_w \omega_c} / 5 z_0$ is the coupling rate. 
Here $z_0 = (5 e^2 / 16 \pi \epsilon_0 M \omega_{\rm{Z}}^2)^{1/3}$ is the equilibrium distance between the ions, $M$ is the ion mass, $e$ is the ion charge, $\epsilon_0$ is the vacuum permittivity, and $\omega_{\rm{Z}}$ is the single ion axial trap frequency. The Hamiltonian~\eqref{eq:fridge_interaction} is valid in the rotating wave approximation when the mode frequencies satisfy the resonance condition $\omega_h = \omega_w + \omega_c$. Away from this resonance condition, energy exchange between modes is suppressed~\cite{Roos2008Nonlinear,Ding2017Cross}. 

The operation of the refrigerator consists of three major steps. First, the Raman beams cool down all modes to the ground state of the trap and starting from there selectively prepare the desired states in the modes corresponding to refrigerator bodies. Then, with the lasers switched off, the resonant trilinear interaction is switched on for some time by tuning the mode frequencies to satisfy the resonance condition such that energy exchange between the modes is allowed. Finally, the mode frequencies are brought back to the initial values and measurement of the resulting state in one of the modes is performed with the help of the Raman beams.

The refrigeration itself occurs during the second step, and to see how it works~\cite{Levy2012Refrigerator}, consider how the resonant interaction Hamiltonian~\eqref{eq:fridge_interaction} redistributes energy between the modes. The work mode ($w$) can only remove one of its excess thermal phonons by creating a hot mode ($h$) phonon and simultaneously annihilating a phonon in the cold mode ($c$). 
Hence the transfer of energy from ($w$) to ($h$) is always accompanied by energy transfer from ($c$) to ($h$). This can result in the cooling of the cold mode when the temperature of the work mode is higher than the hot mode, and energy tends to flow from the former to the latter. This process is balanced by the flow of the energy in opposite direction at some temperature, leading to an equilibrium.

To experimentally demonstrate the equilibrium performance of the refrigerator, all modes are prepared in various thermal states. The system is then allowed to evolve for long interaction times $\tau\gg\xi_{\rm n}^{-1}$, and the mean phonon numbers of each mode is measured to get an estimate of the asymptotic steady state energy. 

%
%


This experiment also allows exploring quantum effects in the operation of heat machines, by comparing the cooling performance starting from a squeezed thermal state of the work mode~\cite{Correa2014refrigerator} to the case where the mode is prepared in a thermal state with the same mean phonon number. The single shot cooling~\cite{Mitchison2015Coherence,Brask2015Small} method, where the interaction is switched off at the right moment such that the evolution halts at a transient state with a lower mean phonon number $\bar{n}_c$ than the long-time average, was also demonstrated. 


\subsection{Vibronic Sampling}
Because of their complexity and the emergent role of quantum nature, molecules are one of the most demanding quantum systems for quantum computers to simulate. For molecular vibronic spectroscopy, it carries the vibrational transitions between nuclear manifolds belonging to two electronic states~\cite{jankowiak2007Vibronic,Huh2015Boson}. Upon an electronic transition, a molecule undergoes structural deformation, vibrational frequency shifts, and rotation of normal modes; within a harmonic approximation to the electronic potential energy surfaces, these are equivalent to the displacement ($\hat{D}$), squeezing ($\hat{S}$) and rotation ($\hat{R}$) operations in quantum optics, respectively. These effects was simplified as a Duschinsky linear relation \cite{duschinsky:1937}. Doktorov et al.~\cite{doktorov1977Dynamical} decomposed the relation in terms of the following operators, 
\begin{align}
\hat{U}_{\mathrm{Dok}}=\hat{D}_N(\boldsymbol{\alpha})\hat{S}^{\dagger}_N(\boldsymbol{\zeta'}) \hat{R}_N(\boldsymbol{\theta})\hat{S}_N(\boldsymbol{\zeta}),
\label{eq:Doktorov3}
\end{align}
where $\hat{D}_N,\hat{S}_N$ and $\hat{R}_N$ are the $N$-mode operators of displacement, squeezing and rotation \cite{Ma1990}, $\boldsymbol{\alpha}$ is a (dimensionless) molecular displacement vector, $\boldsymbol{\zeta}$ and $\boldsymbol{\zeta'}$ are diagonal matrices of the squeezing parameters, and $\boldsymbol{\theta}$ is a vector of rotation angles.

The process of molecular vibronic spectroscopy can be understood as a modified boson sampling with Gaussian input states such as thermal and squeezed vacuum states. The Gaussian boson sampling, which is classified as classically hard problem in the computational complexity perspective~\cite{Lund2014Boson,rahimi2015What}, requires additional quantum optical operations on top of the beam splitting and phase shifting operations for standard boson sampling.

Using a trapped ion, the quantum simulation of molecular vibronic spectroscopy with a particular example of photoelectron spectroscopy of sulfur dioxide  (SO$_{2}$)~\cite{nimlos1986,Lee2009general} is performed. The two vibrational modes of the molecule are mapped to the two radial phonon modes of an ion. Then, the molecular spectroscopy is simulated through the following procedure: (i) the ion is first initialized in the motional ground state, (ii) the quantum optical operations of Doktorov \cite{doktorov1977Dynamical} are sequentially applied, and (iii) finally, the vibronic spectrum is then constructed using the collective projection measurements on the transformed state. 

After preparing the motional ground-state by Doppler cooling and resolved sideband cooling methods~\cite{monroe1995resolved,roos1999quantum}, Doktorov operations are performed as the required displacement, squeezing and rotation operations with the conversion of molecular parameters to the corresponding device parameters as $\boldsymbol{\alpha}, \boldsymbol{\zeta'}, \boldsymbol{U}$ and $\boldsymbol{\zeta}$, respectively. The molecular parameters can be obtained via conventional quantum chemical calculations with available program packages (e.g., Ref.~\cite{Frisch2016Gaussian}). 

As discussed in the previous sections, the quantum optical operations (displacement $\hat{D}$, squeezing $\hat{S}$ and rotation $\hat{R}$) are implemented by Raman laser beams. The spectrum at zero Kelvin from the output measurements of the trapped-ion simulator, reconstructed from the ground to excited states transition intensities, are aligned according to the transition frequencies. Finally, the collective quantum-projection measurement of the final state $\mathrm{\ket{n_X,n_Y}}$ is performed \cite{Um16Phonon,Zhang2018Experimental}.  

\begin{figure}[tb]
\centering
\includegraphics[width=\columnwidth]{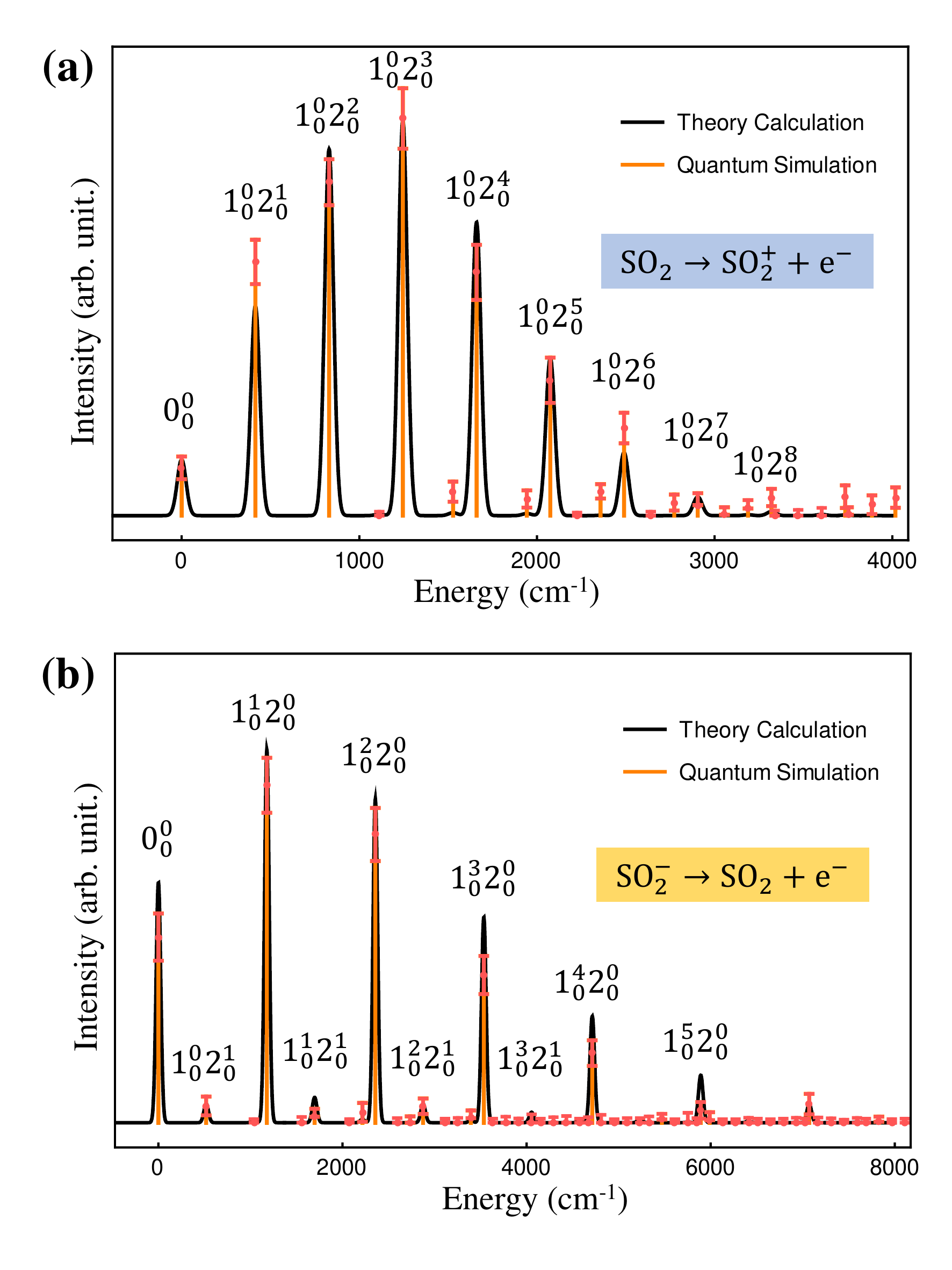}
\caption{\label{fig:SO2}
The simulation result of photoelectron spectra of SO$_2$ and SO$_2^-$. The two vibrational frequencies of harmonic potential for SO$_2^+$, SO$_2$ and SO$_2^-$ are (1112.7, 415.0), (1178.4, 518.9) and (989.5, 451.4) cm$^{-1}$, respectively. (a) The displacement vector $\boldsymbol{\alpha}$ is (-0.026, 1.716); rotation angle $\theta$ is $0.189$ and (b) The displacement vector $\boldsymbol{\alpha}$ is (1.360, -0.264); rotation angle $\theta$ is $0.065$. The results of theoretical calculations are intentionally broadened by convolution with a Gaussian function having a width of 50 cm$^{-1}$ \cite{Lee2009general} for the comparison. Here $N^{i}_{0}$ denotes the $i$-phonon excitation on $N$-th mode from the vibrational ground state $\ket{0}$, and accordingly, $0^{0}_{0}$ located at the off-set energy $\omega_{0-0}=0$. Figure adapted from Ref.~\cite{shen2018quantum}.}
\end{figure}

The photoelectron spectra obtained for SO$_{2}\rightarrow~$SO$_{2}^{+}$ and SO$_{2}^{-}\rightarrow~$SO$_{2}$ are shown in Fig.~\ref{fig:SO2}, which show good agreement with classical calculations. In Fig.~\ref{fig:SO2}(a), the photoelectron spectrum of SO$_2$ is dominated by the transitions originated from the displacement of the second mode, due to the significantly large displacement of the second mode: $\boldsymbol{\alpha} = (-0.026, 1.716)$. The photoelectron spectrum of SO$_2^-$ in Fig.~\ref{fig:SO2}(b) shows tiny but evident effects of the combination of the two modes regardless of the dominant contribution of the first mode ($\boldsymbol{\alpha} =(1.360, -0.264)$). The observation of the band combination indicates reliable performance of the trapped-ion simulation. 

\section{VII. Conclusion and Outlook}
In conclusion, motional degrees of freedom of trapped ions are valuable resources that give access to the large Hilbert space encoded in bosonic modes. Ultimately, we do not expect any fundamental problems of increasing the number of vibrational modes, which can be manipulated in quantum regime, to the order of hundreds with a few phonons per mode in the transverse direction. In a few example discussed above this already leads to various applications in quantum simulations, quantum thermodynamics and even quantum computations. In addition, the quantum simulation that involves phonon states, or combinantion of the phonon and internal states of the ions  \cite{casanova2010deep,mezzacapo2012digital,pedernales2015quantum,Toyoda2013Experimental,lv2018quantum} have addressed various physics problems, including even the problems of relativistic quantum mechanics \cite{Lamata2007Dirac,Gerritsma2010Quantum}, quantum field theory \cite{casanova2011quantum,xiang2018experimental} and others. 

However, several challenges in control of the motional states, both technical and conceptual, currently hinder experimental progress in this area. First limitation in the ion trap experiments comes from detection of motional states, especially when measurement and feedback protocols are employed. Typically, the state of motion in trapped ion experiment is detected by coupling the motional and internal states of the ion, and subsequent fluorescent detection of the ion internal state. Unfortunately during the latter procedure, scattering of thousands or even millions of photons occur when the internal state is a ``bright state''. Most of the time recoil destroys the motional state of all the modes and the experiment has to be restarted. However, it is sometimes possible to harness the recoil provided by the scattered photons to implement feedback protocols ~\cite{deneeve2020error}. In addition, even though the motional state is characterised by continuous variables that describe the position and momentum of the ions, the information extracted though the measurement are mapped to internal states of the ions and corresponds to discrete variables \cite{ohira2019phonon}.  Large and efficient collections of scattered photons can also enable us to perform the repetitive detection of phonon number states before the number states are perturbed \cite{An2015Experimental,Um16Phonon}.   

Anomalous heating and dephasing of the modes typically limit coherence time of harmonic oscillator for the $|0\rangle + |1\rangle$ superpositions to around 10 ms, with dephasing being the dominant decoherence mechanism in most of the experiments. As the impact of both mechanisms increase with the phonon number, this limits the Hilbert space dimension that one can experimentally control, and the number of useful quantum operations that can be performed. Using rf source with stable amplitude for driving the trap electrodes can help to mitigate dephasing~\cite{Monroe2006rf_stability}. Cleaning of the trap before assembly~\cite{nist2012trap_cleaning} and placing the trap into a cryogenic environment~\cite{monroe2006heating} was shown to reduce heating of the modes. In addition, the effects of heating are typically significantly lower for non-center-of-mass modes. Furthermore, the fact that fluctuation of the radial trapping potential affects some modes in a similar manner, with induced errors cancel each other when these modes are manipulated together, gives rise to an extended coherence time compared to that of the individual modes~\cite{Ding2018Trilinear,tamura2020quantum}.   

In the future, as the number of simultaneously controllable modes increases, more advanced applications will be possible. These include universal quantum computation with continuous variables, various machine learning algorithms~\cite{HKLau2017quantum}, such as matrix inversion, principal component analysis, and implementations of support vector machines~\cite{Havlicek2019SVM}. By exploiting a greater number of modes, quantum simulations stand to benefit as well, which includes, for instance, simulations of molecular vibronic spectra and quantum simulation of polaron together with internal states. Ultimately, increase in the number of modes that can be controlled simultaneously may eventually lead to Hilbert spaces whose dimensions exceed the simulation ability of classical computers.  

\providecommand{\newblock}{}


\end{document}